\documentstyle[aps,prb,epsf]{revtex}

\begin{document}

\renewcommand\baselinestretch{1.3}
\large\normalsize

\title{Jordan--Wigner fermionization
       \protect\\
       for spin--$\frac{1}{2}$ systems in two dimensions:
       \protect\\
       A brief review}

\author{Oleg Derzhko\\
{\small {Institute for Condensed Matter Physics}}\\
{\small {1 Svientsitskii Street, L'viv--11, 79011, Ukraine}}\\
{\small {and}}\\
{\small {Chair of Theoretical Physics,
Ivan Franko National University of L'viv}}\\
{\small {12 Drahomanov Street, L'viv--5, 79005, Ukraine}}}

\date{\today}

\maketitle

\begin{abstract}
We review the papers on the Jordan--Wigner transformation
in two dimensions
to comment on a possibility
of examining the statistical mechanics properties of
two--dimensional spin--$\frac{1}{2}$ models.
We discuss in some detail
the two--dimensional
spin--$\frac{1}{2}$
isotropic $XY$ and Heisenberg models.
\end{abstract}

\vspace{5mm}


\vspace{5mm}


\vspace{5mm}

\noindent
{\bf {Postal address:}}\\
Dr. Oleg Derzhko\\
Institute for Condensed Matter Physics\\
1 Svientsitskii Street, L'viv--11, 79011, Ukraine\\
tel/fax: (0322) 76 19 78\\
email: derzhko@icmp.lviv.ua\\


\renewcommand\baselinestretch{1.45}
\large\normalsize

\section{Introductory remarks}

A mapping of the spin--$\frac{1}{2}$ operators
onto Fermi operators by means of the Jordan--Wigner transformation
was used by Lieb, Schultz and Mattis \cite{001}
to introduce the exactly solvable
one--dimensional spin--$\frac{1}{2}$ $XY$ model.
Later the famous Onsager's solution
for the two--dimensional Ising model
was reproduced by using the Jordan--Wigner transformation
for the transfer matrix of that model \cite{002}.
The Jordan--Wigner transformation is the essential ingredient
of the studies of the statistical mechanics properties
of quantum spin chains \cite{003}.
Much effort has been devoted
to generalize the fermionization procedure
for two \cite{004,005,006,007,008,009,010,011,012,013,014}
and three \cite{015,016} dimensions.
In the present paper
we review the Jordan--Wigner transformation
in two dimensions as well as some existing applications
of this mapping
for the spin system theory.

We shall consider a spin model consisting of $N=N_x N_y$
($N_x\to\infty$, $N_y\to\infty$)
spins $\frac{1}{2}$ on a square lattice
of the size $L_x L_y$
($L_x\to\infty$, $L_y\to\infty$)
governed by the Heisenberg Hamiltonian
\begin{eqnarray}
\label{001}
H=\sum_{\langle{\bf{i}},{\bf{j}}\rangle}
J {\bf{s}}_{\bf{i}}\cdot{\bf{s}}_{\bf{j}}
+\sum_{\bf i}
h s_{\bf{i}}^z
\nonumber\\
=\sum_{i=0}^{\infty}\sum_{j=0}^{\infty}
J
\left({\bf{s}}_{i,j}\cdot{\bf{s}}_{i+1,j}
+{\bf{s}}_{i,j}\cdot{\bf{s}}_{i,j+1}\right)
+\sum_{i=0}^{\infty}\sum_{j=0}^{\infty}
h
s_{i,j}^z.
\end{eqnarray}
Here $\langle{\bf{i}},{\bf{j}}\rangle$
denotes all different nearest neighbouring sites
at the square lattice,
$J$ is the exchange interaction between the neighbouring sites,
$h$ is the external field.
The isotropic Heisenberg interaction
consists of the isotropic $XY$ part and the Ising part
\begin{eqnarray}
\label{002}
{\bf{s}}_{\bf{i}}\cdot{\bf{s}}_{\bf{j}}
=\left(s^x_{\bf{i}}s^x_{\bf{j}}+s^y_{\bf{i}}s^y_{\bf{j}}\right)
+s^z_{\bf{i}}s^z_{\bf{j}}
\nonumber\\
=\frac{1}{2}
\left(s^+_{\bf{i}}s^-_{\bf{j}}+s^-_{\bf{i}}s^+_{\bf{j}}\right)
+\left(s^+_{\bf{i}}s^-_{\bf{i}}-\frac{1}{2}\right)
\left(s^+_{\bf{j}}s^-_{\bf{j}}-\frac{1}{2}\right),
\end{eqnarray}
where we have introduced
the spin raising and lowering operators
$s^{\pm}=s^x\pm {\mbox{i}}s^y$
and
$s^z=s^+s^--\frac{1}{2}$.
It is important to note that the operators $s^+$, $s^-$
obey the Fermi commutation rules at the same site
\begin{eqnarray}
\label{003}
\left\{s^-_{{\bf{i}}},s^+_{{\bf{i}}}\right\}=1,
\;\;\;
\left\{s^+_{{\bf{i}}},s^+_{{\bf{i}}}\right\}
=\left\{s^-_{{\bf{i}}},s^-_{{\bf{i}}}\right\}
=0
\end{eqnarray}
and the Bose commutation rules at different sites
${\bf{i}}\ne{\bf{j}}$
\begin{eqnarray}
\label{004}
\left[s^-_{{\bf{i}}},s^+_{{\bf{j}}}\right]
=\left[s^+_{{\bf{i}}},s^+_{{\bf{j}}}\right]
=\left[s^-_{{\bf{i}}},s^-_{{\bf{j}}}\right]
=0.
\end{eqnarray}
The aim of the Jordan--Wigner trick 
is to transform the spin variables into pure fermion variables.

We start the paper 
by giving a short reminder
of the Jordan--Wigner transformation in one dimension
(Section II).
Then we discuss the extensions for two dimensions
suggested by M. Azzouz \cite{008}
(Section III),
Y. R. Wang \cite{005}
(Section IV),
and E. Fradkin \cite{004}
(Section V).
The mean--field--like treatment of the Jordan--Wigner fermions
for the isotropic $XY$ model
is discussed in detail
in Section VI.
The consideration of a model with Ising term
in fermionic language is given separately
in Section VII.
Finally, we summarize some of the results obtained
using this approximate approach
and comment on a comparison with the results
derived using other methods
(Section VIII).

\section{The Jordan--Wigner transformation in one dimension}

With the help of the Jordan--Wigner transformation
we introduce instead of the operators $s^+$, $s^-$
satisfying (\ref{003}), (\ref{004})
the operators $c^+$, $c$
satisfying the Fermi commutation rules
(both at the same and different sites)
in terms of which
the Hamiltonian of the isotropic $XY$ chain
is a bilinear form
and the Ising interaction
yields the products of four Fermi operators.
Explicitly the Jordan--Wigner transformation reads
\begin{eqnarray}
\label{005}
c_n^+={\mbox{e}}^{{\mbox{i}}\alpha_n}s_n^+,
\;\;\;
c_n={\mbox{e}}^{-{\mbox{i}}\alpha_n}s_n^-,
\;\;\;
s_n^+={\mbox{e}}^{-{\mbox{i}}\alpha_n}c_n^+,
\;\;\;
s_n^-={\mbox{e}}^{{\mbox{i}}\alpha_n}c_n,
\nonumber\\
\alpha_n=\pi\sum_{j=0}^{n-1}n_j,
\;\;\;
n_j=c_j^+c_j.
\end{eqnarray}
The signs in the exponents
and the order of the multipliers
(first line in (\ref{005}))
are not important.
To get the operator $c_n$ ($c_n^+$)
we must multiply $s_n^-$ ($s_n^+$)
by the exponent
containing a sum of $n_j$ at all previous sites $0\le j\le n-1$
as can be seen from (\ref{005})
and is shown symbolically in Fig. 1.

\begin{figure}
\epsfysize=20mm
\epsfclipon
\centerline{\epsffile{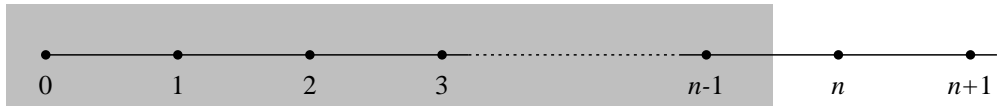}}
\caption[ ]
{\small Towards the Jordan--Wigner transformation in one dimension.}
\label{fig1}
\end{figure}

To demonstrate how the transformation (\ref{005}) works
we need the following equations for Fermi operators
\begin{eqnarray}
\label{006}
c_j^+{\mbox{e}}^{{\mbox{i}}\pi c_j^+c_j}=c_j^+,
\;\;\;
{\mbox{e}}^{{\mbox{i}}\pi c_j^+c_j}c_j^+=-c_j^+,
\;\;\;
c_j{\mbox{e}}^{{\mbox{i}}\pi c_j^+c_j}=-c_j,
\;\;\;
{\mbox{e}}^{{\mbox{i}}\pi c_j^+c_j}c_j=c_j,
\end{eqnarray}
which can be easily obtained since
$$
{\mbox{e}}^{{\mbox{i}}\pi c_j^+c_j}
=1+{\mbox{i}}\pi c_j^+c_j
+\frac{1}{2!}\left({\mbox{i}}\pi\right)^2 c_j^+c_j
+\ldots
=1-c_j^+c_j+{\mbox{e}}^{{\mbox{i}}\pi}c_j^+c_j
=1-2c_j^+c_j.
$$
From (\ref{006}) one finds that
\begin{eqnarray}
\label{007}
{\mbox{e}}^{{\mbox{i}}\pi c_n^+c_n}c_l
=c_l{\mbox{e}}^{{\mbox{i}}\pi c_n^+c_n},
\;\;\;
{\mbox{e}}^{{\mbox{i}}\pi c_n^+c_n}c_l^+
=c_l^+{\mbox{e}}^{{\mbox{i}}\pi c_n^+c_n},
\end{eqnarray}
if $n\ne l$,
but
\begin{eqnarray}
\label{008}
{\mbox{e}}^{{\mbox{i}}\pi c_l^+c_l}c_l
=-c_l{\mbox{e}}^{{\mbox{i}}\pi c_l^+c_l},
\;\;\;
{\mbox{e}}^{{\mbox{i}}\pi c_l^+c_l}c_l^+
=-c_l^+{\mbox{e}}^{{\mbox{i}}\pi c_l^+c_l}.
\end{eqnarray}
Besides,
\begin{eqnarray}
\label{009}
{\mbox{e}}^{2{\mbox{i}}\pi c_l^+c_l}=1.
\end{eqnarray}

Let us show
that the commutation rules
for the spin raising and lowering operators
$s^+$, $s^-$ (\ref{005})
are given by (\ref{003}), (\ref{004})
if $c^+$, $c$ are Fermi operators.
At the same site we find
\begin{eqnarray}
\label{010}
s_n^-s_n^+
=c_n{\mbox{e}}^{{\mbox{i}}\alpha_n}{\mbox{e}}^{-{\mbox{i}}\alpha_n}c_n^+
=c_nc_n^+,
\;\;\;
s_n^+s_n^-=c_n^+c_n,
\;\;\;
s_n^+s_n^+=c_n^+c_n^+=0,
\;\;\;
s_n^-s_n^-=c_nc_n=0,
\end{eqnarray}
and as a result Eq. (\ref{003}) becomes evident.
At different sites
$n$ and $n+m$ (without any loss of generality $m>0$)
we have
\begin{eqnarray}
\label{011}
s_n^-s_{n+m}^+
=c_n{\mbox{e}}^{{\mbox{i}}\pi \sum_{l=n}^{n+m-1}c_l^+c_l}c_{n+m}^+
=c_nc_{n+m}^+{\mbox{e}}^{{\mbox{i}}\pi \sum_{l=n}^{n+m-1}c_l^+c_l},
\nonumber\\
s_{n+m}^+s_n^-
=c_{n+m}^+{\mbox{e}}^{{\mbox{i}}\pi \sum_{l=n}^{n+m-1}c_l^+c_l}c_n
=-c_{n+m}^+c_n{\mbox{e}}^{{\mbox{i}}\pi \sum_{l=n}^{n+m-1}c_l^+c_l},
\end{eqnarray}
etc. which immediately yields Eq. (\ref{004}).

Let us write down the transformed Hamiltonian.
Besides
$s_j^+s_j^-=c_j^+c_j$
(\ref{010})
we have
\begin{eqnarray}
\label{012}
s_j^+s_{j+1}^-
=c_j^+{\mbox{e}}^{{\mbox{i}}\pi c_j^+c_j}c_{j+1}
=c_j^+c_{j+1},
\;\;\;
s_j^-s_{j+1}^+
=c_j{\mbox{e}}^{{\mbox{i}}\pi c_j^+c_j}c_{j+1}^+
=-c_jc_{j+1}^+,
\end{eqnarray}
and as a result
the one--dimensional spin--$\frac{1}{2}$ Heisenberg Hamiltonian
(\ref{001}), (\ref{002})
becomes
\begin{eqnarray}
\label{013}
H=\sum_j
\left(
\frac{1}{2}J\left(c_j^+c_{j+1}-c_jc_{j+1}^+\right)
+J\left(c_j^+c_j-\frac{1}{2}\right)
\left(c_{j+1}^+c_{j+1}-\frac{1}{2}\right)
\right)
+\sum_j
h \left(c_j^+c_j-\frac{1}{2}\right).
\end{eqnarray}

It is clear now to what extent the Jordan--Wigner transformation
simplifies further statistical mechanics calculations.
A nontrivial in spin language isotropic $XY$ chain
transforms into tight--binding spinless fermions
and further rigorous treatment becomes possible.
For the anisotropic $XY$ chain
the operators $c_j^+c_{j+1}^+$, $c_jc_{j+1}$
enter Eq. (\ref{013}) besides,
and therefore the Bogolyubov transformation is required in addition.
The Ising term leads to an interaction
between the Jordan--Wigner fermions,
however, the low--energy properties
may be analysed using the bosonization techniques \cite{003}.
While interesting only in the low--energy physics
of $s>\frac{1}{2}$ spin chains
one may represent the spin--$s$ operators
as a sum of $2s$ spin--$\frac{1}{2}$ operators
and then,
making use of the Jordan--Wigner representation
for the latter operators,
proceed in fermionic language.

\section{The Jordan--Wigner transformation in two dimensions
(M. Azzouz, 1993)}

Consider the spin model (\ref{001}) on a square lattice (Fig. 2).
%
\begin{figure}
\epsfysize=100mm
\epsfclipon
\centerline{\epsffile{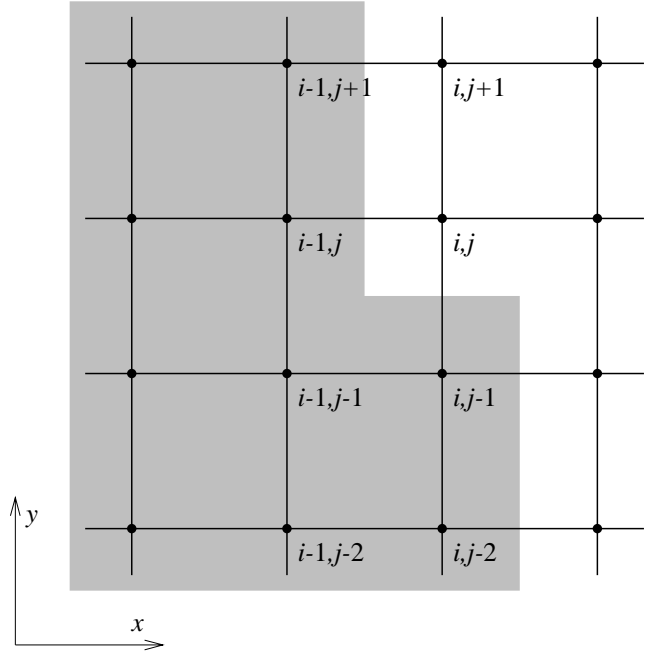}}
\caption[ ]
{\small Towards the Jordan--Wigner transformation in two dimensions
\cite{008}.}
\label{fig2}
\end{figure}
Two coordinates $i$ and $j$ are taken at the $x$ and $y$ axes,
respectively,
to specify a given site.

M. Azzouz defined \cite{008}
the extended Jordan--Wigner transformation as
\begin{eqnarray}
\label{014}
s_{i,j}^-
=c_{i,j}{\mbox{e}}^{{\mbox{i}}\alpha_{i,j}}=
{\mbox{e}}^{{\mbox{i}}\alpha_{i,j}}c_{i,j},
\;\;\;
s_{i,j}^+
=c_{i,j}^+{\mbox{e}}^{-{\mbox{i}}\alpha_{i,j}}=
{\mbox{e}}^{-{\mbox{i}}\alpha_{i,j}}c_{i,j}^+,
\nonumber\\
\alpha_{i,j}
=\pi\left(\sum_{d=0}^{i-1}\sum_{f=0}^{\infty}n_{d,f}
+\sum_{f=0}^{j-1}n_{i,f}\right),
\;\;\;
n_{d,f}=c^+_{d,f}c_{d,f}
\end{eqnarray}
(compare with Eqs. (\ref{005})).
The signs in the exponents in (\ref{014})
(and the order of the multipliers
in the first line in (\ref{014}))
are not important.
Let us show that the introduced transformation (\ref{014})
enables one to construct
a fermion representation
for two--dimensional spin--$\frac{1}{2}$ models.

We start with the commutation rules.
At the same site one has
\begin{eqnarray}
\label{015}
s_{q,p}^-s_{q,p}^+
=c_{q,p}{\mbox{e}}^{{\mbox{i}}\alpha_{q,p}}
{\mbox{e}}^{-{\mbox{i}}\alpha_{q,p}} c_{q,p}^+
=c_{q,p}c_{q,p}^+,
\;\;\;
s_{q,p}^+s_{q,p}^-
=c_{q,p}^+c_{q,p},
\;\;\;
s_{q,p}^+s_{q,p}^+
=c_{q,p}^+c_{q,p}^+,
\;\;\;
s_{q,p}^-s_{q,p}^-
=c_{q,p}c_{q,p},
\end{eqnarray}
and the Fermi type commutation rules remain unchanged.
To illustrate
how transformation (\ref{014}) works
in the case of different sites
let us consider, for example, two sites
$q,p$ and $q,p+m$, $m>0$.
Then, similarly to Eq. (\ref{011}) one finds
\begin{eqnarray}
\label{016}
s_{q,p}^-s_{q,p+m}^+
=c_{q,p}
{\mbox{e}}^{{\mbox{i}}\pi\sum_{f=p}^{p+m-1}n_{q,f}}
c_{q,p+m}^+
=c_{q,p}c_{q,p+m}^+
{\mbox{e}}^{{\mbox{i}}\pi\sum_{f=p}^{p+m-1}n_{q,f}},
\nonumber\\
s_{q,p+m}^+s_{q,p}^-
=c_{q,p+m}^+
{\mbox{e}}^{{\mbox{i}}\pi\sum_{f=p}^{p+m-1}n_{q,f}}
c_{q,p}
=-c_{q,p+m}^+c_{q,p}
{\mbox{e}}^{{\mbox{i}}\pi\sum_{f=p}^{p+m-1}n_{q,f}}
\end{eqnarray}
(we have used an analogue of Eqs. (\ref{007}), (\ref{008}) 
in two dimensions)
and hence the operators 
$s_{q,p}^-$ and $s_{q,p+m}^+$
commute if the operators 
$c_{q,p}$ and $c_{q,p+m}^+$
anticommute.
Following the same reasoning 
one can check the rest of the commutation rules.

Consider further the transformed spin Hamiltonian.
Let us treat somewhat more general nearest neighbour interactions
than the one in Eq. (\ref{001}).
Namely, we assume different values of interaction
in different directions on a square lattice
as shown in Fig. 3.
%
\begin{figure}
\epsfysize=100mm
\epsfclipon
\centerline{\epsffile{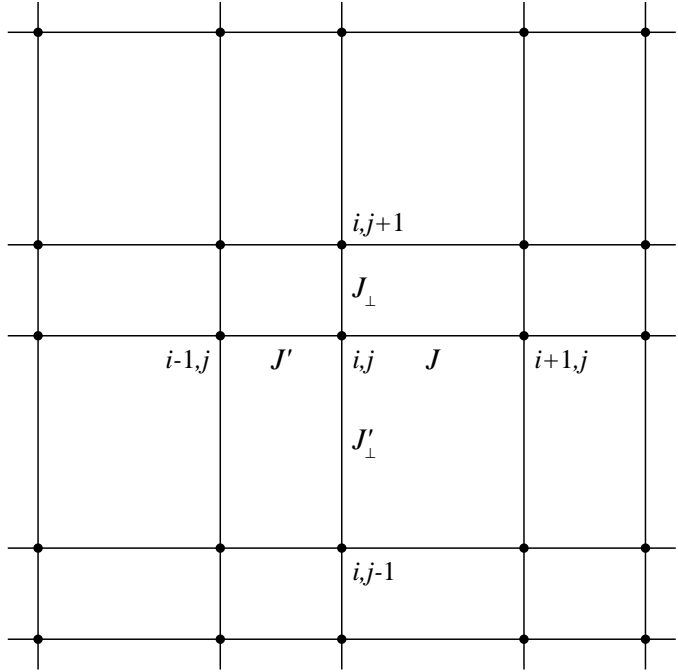}}
\caption[ ]
{\small Nearest neighbour interactions on a square lattice.}
\label{fig3}
\end{figure}
(Evidently, 
we can perform the fermionization presented below 
for a completely nonuniform model 
characterized by a set of intersite interactions
$\left\{\ldots, J_{i,j;i+1,j}, \ldots;
\ldots, J_{i,j;i,j+1}, \ldots\right\}$.)
If
$J=J^{\prime}=J_{\perp}=J_{\perp}^{\prime}$
one faces the uniform square lattice.
To examine the effects of interchain interactions
in quasi--one--dimensional systems
one may consider the case
$J_{\perp},\;J_{\perp}^{\prime}\ll J,\;J^{\prime}$
(with $J$ not equal to $J^{\prime}$
if the dimerised chain is considered).
If
$J_{\perp}^{\prime}\ll J,J^{\prime},J_{\perp}$
one has a model of interacting two--leg ladders.
In the limiting case of 
$J_{\perp}^{\prime}=0$
(the noninteracting two--leg ladders)
the model may be reduced to a one--dimensional system
with interactions extending over the nearest sites.

We begin with the isotropic $XY$ interaction
(see Fig. 4).
%
\begin{figure}
\epsfysize=80mm
\epsfclipon
\centerline{\epsffile{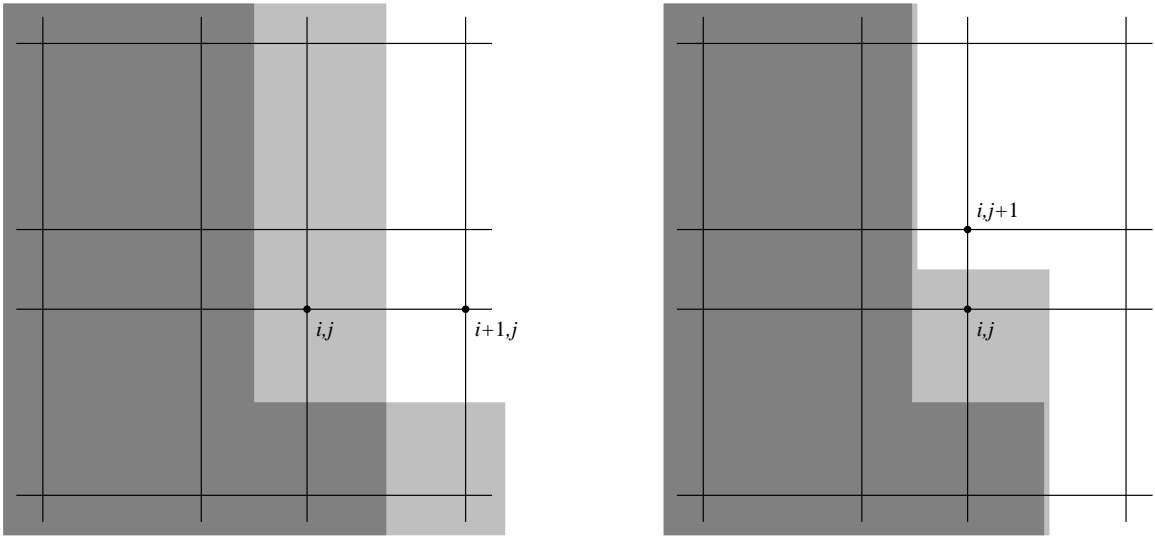}}
\caption[ ]
{\small Towards the fermionization of $H_{XY}$.}
\label{fig4}
\end{figure}
Inserting (\ref{014}) into (\ref{001}), (\ref{002})
one finds
\begin{eqnarray}
\label{017}
H_{XY}
=\frac{1}{2}\sum_{i=0}^{\infty}\sum_{j=0}^{\infty}
\left(
J_{i,j;i+1,j}
\left(s_{i,j}^-s_{i+1,j}^++s_{i,j}^+s_{i+1,j}^-\right)
+
J_{i,j;i,j+1}
\left(s_{i,j}^-s_{i,j+1}^++s_{i,j}^+s_{i,j+1}^-\right)
\right)
\nonumber\\
=\frac{1}{2}\sum_{i=0}^{\infty}\sum_{j=0}^{\infty}
\left(
J_{i,j;i+1,j}
\left(c_{i,j}
{\mbox{e}}^{-{\mbox{i}}\pi\left(\sum_{f=j}^{\infty}n_{i,f}
+\sum_{f=0}^{j-1}n_{i+1,f}\right)}
c_{i+1,j}^+
+c_{i,j}^+
{\mbox{e}}^{{\mbox{i}}\pi\left(\sum_{f=j}^{\infty}n_{i,f}
+\sum_{f=0}^{j-1}n_{i+1,f}\right)}
c_{i+1,j}\right)
\right.
\nonumber\\
\left.
+
J_{i,j;i,j+1}
\left(c_{i,j}
{\mbox{e}}^{-{\mbox{i}}\pi n_{i,j}}
c_{i,j+1}^+
+c_{i,j}^+
{\mbox{e}}^{{\mbox{i}}\pi n_{i,j}}
c_{i,j+1}\right)
\right)
\nonumber\\
=\frac{1}{2}\sum_{i=0}^{\infty}\sum_{j=0}^{\infty}
\left(
J_{i,j;i+1,j}
\left(-c_{i,j}
{\mbox{e}}^{-{\mbox{i}}\pi\left(\sum_{f=j+1}^{\infty}n_{i,f}
+\sum_{f=0}^{j-1}n_{i+1,f}\right)}
c_{i+1,j}^+
+c_{i,j}^+
{\mbox{e}}^{{\mbox{i}}\pi\left(\sum_{f=j+1}^{\infty}n_{i,f}
+\sum_{f=0}^{j-1}n_{i+1,f}\right)}
c_{i+1,j}\right)
\right.
\nonumber\\
\left.
+
J_{i,j;i,j+1}
\left(-c_{i,j}c_{i,j+1}^++c_{i,j}^+c_{i,j+1}\right)
\right)
\end{eqnarray}
(to get the last equality
we have used an analogue of Eq. (\ref{006}) in two dimensions).
After introducing the notations
\begin{eqnarray}
\label{018}
\phi_{i,i+1}(j)=
\pi\left(\sum_{f=j}^{\infty}n_{i,f}+\sum_{f=0}^{j-1}n_{i+1,f}\right),
\nonumber\\
\tilde{\phi}_{i,i+1}(j)=
\pi\left(\sum_{f=j+1}^{\infty}n_{i,f}+\sum_{f=0}^{j-1}n_{i+1,f}\right),
\nonumber\\
\varphi_{j,j+1}(i)=\pi n_{i,j}
\end{eqnarray}
the Hamiltonian (\ref{017}) becomes as follows
\begin{eqnarray}
\label{019}
H_{XY}
=\frac{1}{2}\sum_{i=0}^{\infty}\sum_{j=0}^{\infty}
\left(
J_{i,j;i+1,j}
\left(c_{i,j}
{\mbox{e}}^{-{\mbox{i}}\phi_{i,i+1}(j)}
c_{i+1,j}^+
+c_{i,j}^+
{\mbox{e}}^{{\mbox{i}}\phi_{i,i+1}(j)}
c_{i+1,j}\right)
\right.
\nonumber\\
\left.
+
J_{i,j;i,j+1}
\left(c_{i,j}
{\mbox{e}}^{-{\mbox{i}}\varphi_{j,j+1}(i)}
c_{i,j+1}^+
+c_{i,j}^+
{\mbox{e}}^{{\mbox{i}}\varphi_{j,j+1}(i)}
c_{i,j+1}\right)
\right)
\nonumber\\
=\frac{1}{2}\sum_{i=0}^{\infty}\sum_{j=0}^{\infty}
\left(
J_{i,j;i+1,j}
\left(-c_{i,j}
{\mbox{e}}^{-{\mbox{i}}\tilde{\phi}_{i,i+1}(j)}
c_{i+1,j}^+
+c_{i,j}^+
{\mbox{e}}^{{\mbox{i}}\tilde{\phi}_{i,i+1}(j)}
c_{i+1,j}\right)
\right.
\nonumber\\
\left.
+
J_{i,j;i,j+1}
\left(-c_{i,j}c_{i,j+1}^++c_{i,j}^+c_{i,j+1}\right)
\right).
\end{eqnarray}
Eq. (\ref{019}) can be viewed
as the Hamiltonian
of a two--dimensional tight--binding--like spinless fermions
with the hopping amplitudes
\begin{eqnarray}
\label{020}
\mp\frac{1}{2}J_{i,j;i+1,j}
{\mbox{e}}^{\mp {\mbox{i}}\tilde{\phi}_{i,i+1}(j)}
\end{eqnarray}
in the $x$ direction
and
\begin{eqnarray}
\label{021}
\mp\frac{1}{2}J_{i,j;i,j+1}
\end{eqnarray}
in the $y$ direction.
Those hoppings depend in a complicated way
on a configuration of the `intermediate' sites.
Their complexity explains
how the isotropic $XY$ model becomes difficult to examine
in two dimensions
in comparison with an obvious analysis
in one dimension.

There are no difficulties
in rewriting the Ising interaction in fermionic language
\begin{eqnarray}
\label{022}
H_{Z}
=\sum_{i=0}^\infty\sum_{j=0}^\infty
\left(
J_{i,j;i+1,j}
\left(s_{i,j}^+s_{i,j}^--\frac{1}{2}\right)
\left(s_{i+1,j}^+s_{i+1,j}^--\frac{1}{2}\right)
+J_{i,j;i,j+1}
\left(s_{i,j}^+s_{i,j}^--\frac{1}{2}\right)
\left(s_{i,j+1}^+s_{i,j+1}^--\frac{1}{2}\right)
\right)
\nonumber\\
=\sum_{i=0}^\infty\sum_{j=0}^\infty
\left(
J_{i,j;i+1,j}
\left(c_{i,j}^+c_{i,j}-\frac{1}{2}\right)
\left(c_{i+1,j}^+c_{i+1,j}-\frac{1}{2}\right)
+J_{i,j;i,j+1}
\left(c_{i,j}^+c_{i,j}-\frac{1}{2}\right)
\left(c_{i,j+1}^+c_{i,j+1}-\frac{1}{2}\right)
\right)
\end{eqnarray}
and the interaction with an external field
\begin{eqnarray}
\label{023}
H_{f}
=\sum_{i=0}^\infty\sum_{j=0}^\infty
h
\left(s_{i,j}^+s_{i,j}^--\frac{1}{2}\right)
=\sum_{i=0}^\infty\sum_{j=0}^\infty
h
\left(c_{i,j}^+c_{i,j}-\frac{1}{2}\right).
\end{eqnarray}

Formulas (\ref{019}), (\ref{022}), (\ref{023})
realize the fermionic representation
of the spin--$\frac{1}{2}$ isotropic Heisenberg model
on a square lattice (\ref{001}), (\ref{002}).

\section{The Jordan--Wigner transformation in two dimensions
(Y. R. Wang, 1991)}

Let us turn back to the Jordan--Wigner transformation
in one dimension (\ref{005})
using this case as a guideline
and define a particle--annihilation operator as
\begin{eqnarray}
\label{024}
d_{{\bf{i}}}
={\mbox{e}}^{-{\mbox{i}}\alpha_{{\bf{i}}}}s_{{\bf{i}}}^-,
\;\;\;
\alpha_{{\bf{i}}}=\sum_{{\bf{j}}(\ne{\bf{i}})}
B_{{\bf{i}}{\bf{j}}}n_{{\bf{j}}},
\end{eqnarray}
where $B_{{\bf{i}}{\bf{j}}}$ is the c--number matrix element,
$n_{{\bf{j}}}=d_{{\bf{j}}}^+d_{{\bf{j}}}$.
A particle--creation operator is given by
\begin{eqnarray}
\label{025}
d_{{\bf{i}}}^+
=s_{{\bf{i}}}^+{\mbox{e}}^{{\mbox{i}}\alpha_{{\bf{i}}}}
={\mbox{e}}^{{\mbox{i}}\alpha_{{\bf{i}}}}s_{{\bf{i}}}^+,
\end{eqnarray}
whereas the inverse to Eqs. (\ref{024}), (\ref{025}) formulas read
\begin{eqnarray}
\label{026}
s_{{\bf{i}}}^-
={\mbox{e}}^{{\mbox{i}}\alpha_{{\bf{i}}}}d_{{\bf{i}}}
=d_{{\bf{i}}}{\mbox{e}}^{{\mbox{i}}\alpha_{{\bf{i}}}},
\end{eqnarray}
\begin{eqnarray}
\label{027}
s_{{\bf{i}}}^+
={\mbox{e}}^{-{\mbox{i}}\alpha_{{\bf{i}}}}d_{{\bf{i}}}^+
=d_{{\bf{i}}}^+{\mbox{e}}^{-{\mbox{i}}\alpha_{{\bf{i}}}}.
\end{eqnarray}

We want the introduced operators
$d^+$, $d$
to obey the Fermi type commutation rules.
They are indeed the Fermi operators at the same site
due to Eq. (\ref{003}).
Consider further two different sites
${\bf{i}}\ne{\bf{j}}$.
Assuming that
$d^+$, $d$
are Fermi operators
one finds
\begin{eqnarray}
\label{028}
\left[s_{{\bf{i}}}^+,s_{{\bf{j}}}^-\right]
={\mbox{e}}^{-{\mbox{i}}\alpha_{{\bf{i}}}}
d_{{\bf{i}}}^+
{\mbox{e}}^{{\mbox{i}}\alpha_{{\bf{j}}}}
d_{{\bf{j}}}
-
{\mbox{e}}^{{\mbox{i}}\alpha_{{\bf{j}}}}
d_{{\bf{j}}}
{\mbox{e}}^{-{\mbox{i}}\alpha_{{\bf{i}}}}
d_{{\bf{i}}}^+
\nonumber\\
=
{\mbox{e}}^{-{\mbox{i}}n_{{\bf{j}}}B_{{\bf{ij}}}}
d_{{\bf{i}}}^+
{\mbox{e}}^{{\mbox{i}}n_{{\bf{i}}}B_{{\bf{ji}}}}
d_{{\bf{j}}}
-
{\mbox{e}}^{{\mbox{i}}n_{{\bf{i}}}B_{{\bf{ji}}}}
d_{{\bf{j}}}
{\mbox{e}}^{-{\mbox{i}}n_{{\bf{j}}}B_{{\bf{ij}}}}
d_{{\bf{i}}}^+.
\end{eqnarray}
Since
$$
{\mbox{e}}^{-{\mbox{i}}n_{{\bf{j}}}B_{{\bf{ij}}}}
=1+\left({\mbox{e}}^{-{\mbox{i}}B_{{\bf{ij}}}}-1\right)n_{{\bf{j}}},
\;\;\;
{\mbox{e}}^{{\mbox{i}}n_{{\bf{i}}}B_{{\bf{ji}}}}
=1+\left({\mbox{e}}^{{\mbox{i}}B_{{\bf{ji}}}}-1\right)n_{{\bf{i}}}
$$
one can easily proceed in calculation of the r.h.s. of Eq. (\ref{028})
finally arriving at
\begin{eqnarray}
\label{029}
\left[s_{{\bf{i}}}^+,s_{{\bf{j}}}^-\right]
=\left(1+{\mbox{e}}^{-{\mbox{i}}B_{{\bf{ij}}}}
{\mbox{e}}^{{\mbox{i}}B_{{\bf{ji}}}}\right)
d_{{\bf{i}}}^+d_{{\bf{j}}}.
\end{eqnarray}
The result equals to $0$
(as it should for the commutator of the operators $s^+$ and $s^-$
attached to different sites)
if
\begin{eqnarray}
\label{030}
{\mbox{e}}^{{\mbox{i}}B_{{\bf{ij}}}}
=-{\mbox{e}}^{{\mbox{i}}B_{{\bf{ji}}}}.
\end{eqnarray}
Thus we assume that $B_{{\bf{ij}}}$'s
in (\ref{024}), (\ref{025}), (\ref{026}), (\ref{027})
satisfy relation (\ref{030})
that yields the spin commutation relations for
$s^+$, $s$ (\ref{003}), (\ref{004})
if
$d^+$, $d$ are Fermi operators.

Y. R. Wang suggested \cite{005}
the following choice for $B_{{\bf{ij}}}$.
Consider
two complex numbers
\begin{eqnarray}
\label{031}
\tau_{{\bf{i}}}
=i_x+{\mbox{i}}i_y
\end{eqnarray}
and
\begin{eqnarray}
\label{032}
\tau_{{\bf{j}}}
=j_x+{\mbox{i}}j_y,
\end{eqnarray}
which correspond to the sites
${\bf{i}}=i_x{\bf{n}}_x+i_y{\bf{n}}_y$
and
${\bf{j}}=j_x{\bf{n}}_x+j_y{\bf{n}}_y$,
respectively.
Here ${\bf{n}}_x$ and ${\bf{n}}_y$
are the unit vectors directed along $x$ and $y$ axes,
respectively.
Assume that
\begin{eqnarray}
\label{033}
B_{{\bf{ij}}}
={\mbox{arg}}\left(\tau_{{\bf{j}}}-\tau_{{\bf{i}}}\right).
\end{eqnarray}
Evidently
\begin{eqnarray}
\label{034}
{\mbox{e}}^{{\mbox{i}}B_{{\bf{ji}}}}
={\mbox{e}}^{{\mbox{i}}{\mbox{arg}}
\left(\tau_{{\bf{i}}}-\tau_{{\bf{j}}}\right)}
={\mbox{e}}^{{\mbox{i}}\left({\mbox{arg}}
\left(\tau_{{\bf{j}}}-\tau_{{\bf{i}}}\right)\pm\pi\right)}
=-{\mbox{e}}^{{\mbox{i}}B_{{\bf{ij}}}},
\end{eqnarray}
that is the required condition (\ref{030}).
Since
$\tau_{{\bf{i}}}-\tau_{{\bf{j}}}
=\vert \tau_{{\bf{i}}}-\tau_{{\bf{j}}}\vert
{\mbox{e}}^{{\mbox{i}}{\mbox{arg}}
\left(\tau_{{\bf{i}}}-\tau_{{\bf{j}}}\right)}$
Eq. (\ref{033}) can be rewritten in the form
\begin{eqnarray}
\label{035}
B_{{\bf{ij}}}
={\mbox{Im}}\ln\left(\tau_{{\bf{j}}}-\tau_{{\bf{i}}}\right),
\end{eqnarray}
and hence the introduced transformations
(\ref{024}), (\ref{025}), (\ref{026}), (\ref{027})
contains
\begin{eqnarray}
\label{036}
\alpha_{{\bf{i}}}
=\sum_{{\bf{j}}(\ne{\bf{i}})}
{\mbox{Im}}\ln\left(\tau_{{\bf{j}}}-\tau_{{\bf{i}}}\right)
n_{{\bf{j}}}.
\end{eqnarray}

It is worth noting that the transformation of M. Azzouz (\ref{014})
can also be written as Eqs. (\ref{026}), (\ref{027})
with
\begin{eqnarray}
\label{037}
B_{{\bf{ij}}}
=\pi\left(
\Theta(i_x-j_x)
\left(1-\delta_{i_x,j_x}\right)
+
\delta_{i_x,j_x}
\Theta(i_y-j_y)
\left(1-\delta_{i_y,j_y}\right)
\right)
\end{eqnarray}
where
$\Theta(x)$ is the step function,
and a necessary condition
for having spin to fermion mapping (\ref{030}) fulfilled.

The advantage of the choice of Y. R. Wang
(\ref{036})
becomes clear
when one tries to introduce
an approximate treatment
of the transformed Hamiltonian.
After inserting (\ref{026}), (\ref{027})
into (\ref{001}), (\ref{002})
one finds
\begin{eqnarray}
\label{038}
H=\sum_{\langle {\bf{i}},{\bf{j}} \rangle}
\left(
\frac{1}{2}J_{{\bf{i}},{\bf{j}}}
\left(
d^+_{{\bf{i}}}
{\mbox{e}}^{{\mbox{i}}
\left(\alpha_{{\bf{j}}}-\alpha_{{\bf{i}}}\right)}
d_{{\bf{j}}}
+
d_{{\bf{i}}}
{\mbox{e}}^{{\mbox{i}}
\left(\alpha_{{\bf{i}}}-\alpha_{{\bf{j}}}\right)}
d^+_{{\bf{j}}}
\right)
\right.
\nonumber\\
\left.
+J_{{\bf{i}},{\bf{j}}}
\left(d^+_{{\bf{i}}}d_{{\bf{i}}}-\frac{1}{2}\right)
\left(d^+_{{\bf{j}}}d_{{\bf{j}}}-\frac{1}{2}\right)
\right)
+\sum_{{\bf{i}}}
h\left(d^+_{{\bf{i}}}d_{{\bf{i}}}-\frac{1}{2}\right).
\end{eqnarray}

Note that
\begin{eqnarray}
\label{039}
\alpha_{{\bf{j}}}-\alpha_{{\bf{i}}}
=\int_{{\bf{i}}}^{{\bf{j}}}
{\mbox{d}}{\bf{r}}\cdot{\bf{A}}({\bf{r}})
\end{eqnarray}
where
$
{\bf{A}}({\bf{r}})
=\nabla_{{\bf{r}}}\alpha_{{\bf{r}}}
$
and hence
\begin{eqnarray}
\label{040}
{\bf{A}}({\bf{r}})
=\frac{\partial}{\partial r_x}
\left(
\sum_{{\bf{r}}^{\prime}(\ne{\bf{r}})}
n_{{\bf{r}}^{\prime}}
{\mbox{Im}}\ln
\left(
r^{\prime}_x-r_x
+{\mbox{i}}
\left(
r^{\prime}_y-r_y
\right)
\right)
\right)
{\bf{n}}_x
\nonumber\\
+\frac{\partial}{\partial r_y}
\left(
\sum_{{\bf{r}}^{\prime}(\ne{\bf{r}})}
n_{{\bf{r}}^{\prime}}
{\mbox{Im}}\ln
\left(
r^{\prime}_x-r_x
+{\mbox{i}}
\left(
r^{\prime}_y-r_y
\right)
\right)
\right)
{\bf{n}}_y
\nonumber\\
=\sum_{{\bf{r}}^{\prime}(\ne{\bf{r}})}
n_{{\bf{r}}^{\prime}}
\frac{\left(r^{\prime}_y-r_y\right){\bf{n}}_x
-\left(r^{\prime}_x-r_x\right){\bf{n}}_y}
{\left({\bf{r}}^{\prime}-{\bf{r}}\right)^2}
\nonumber\\
=-\sum_{{\bf{r}}^{\prime}(\ne{\bf{r}})}
n_{{\bf{r}}^{\prime}}
\frac{{\bf{n}}_z\times
\left({\bf{r}}^{\prime}-{\bf{r}}\right)}
{\left({\bf{r}}^{\prime}-{\bf{r}}\right)^2}.
\end{eqnarray}

The crucial approximation to proceed
is to make the change in (\ref{040})
\begin{eqnarray}
\label{041}
n_{{\bf{r}}}\rightarrow
\langle n_{{\bf{r}}}\rangle
=\langle s^z_{{\bf{r}}}\rangle+\frac{1}{2}
\rightarrow \frac{1}{2}.
\end{eqnarray}
Here the angular brackets
denote the thermodynamical canonical average
with the Hamiltonian (\ref{001}).
Thus, we have postulated the mean--field description
of the phase factors in (\ref{038}).
A similar treatment was adopted in Refs. \onlinecite{017,018}.
Apparently, assuming further in (\ref{041})
that $\langle s^z_{{\bf{r}}}\rangle=0$
one should suppose that $h=0$
(although in Refs. \onlinecite{010,011}
the uniform magnetic field
was included into the Hamiltonian).
In principle Eq. (\ref{041}) simplifies the problem drastically 
since one faces a tight--binding spinless fermions on a square lattice.
However, in practice it is hard to proceed
because of nonuniformity of that model. 

The Hamiltonian (\ref{038}), (\ref{039}), (\ref{040}), (\ref{041})
describes
the (charged) spinless fermions
moving in a plane
in an external uniform (classical) magnetic field
which is perpendicular to the plane.
Due to (\ref{041})
$\langle n_{{\bf{r}}^{\prime}}\rangle$
can be taken out from the summation
and in the continuum limit
the vector potential of the field ${\bf{A}}({\bf{r}})$
can be written \cite{007}
in the following form
\begin{eqnarray}
\label{042}
{\bf{A}}({\bf{r}})
=-\langle n_{{\bf{r}}^{\prime}}\rangle
\sum_{{\bf{r}}^{\prime}(\ne{\bf{r}})}
\frac{{\bf{n}}_z\times
\left({\bf{r}}^{\prime}-{\bf{r}}\right)}
{\left({\bf{r}}^{\prime}-{\bf{r}}\right)^2}
\nonumber\\
=-\langle n_{{\bf{r}}^{\prime}}\rangle
\frac{1}{S_0}
\int_{-\frac{L_x}{2}}^{\frac{L_x}{2}}{\mbox{d}}r_x^{\prime}
\int_{-\frac{L_y}{2}}^{\frac{L_y}{2}}{\mbox{d}}r_y^{\prime}
\frac{{\bf{n}}_z\times
\left({\bf{r}}^{\prime}-{\bf{r}}\right)}
{\left({\bf{r}}^{\prime}-{\bf{r}}\right)^2}
\nonumber\\
=\ldots=\langle n_{{\bf{r}}^{\prime}}\rangle
\frac{\pi}{S_0}
{\bf{n}}_z\times{\bf{r}},
\end{eqnarray}
where $S_0$ is the area of the elementary plaquette in the plane
(see Fig. 5) 
%
\begin{figure}
\vspace{-60mm}
\epsfysize=160mm
\epsfclipon
\centerline{\epsffile{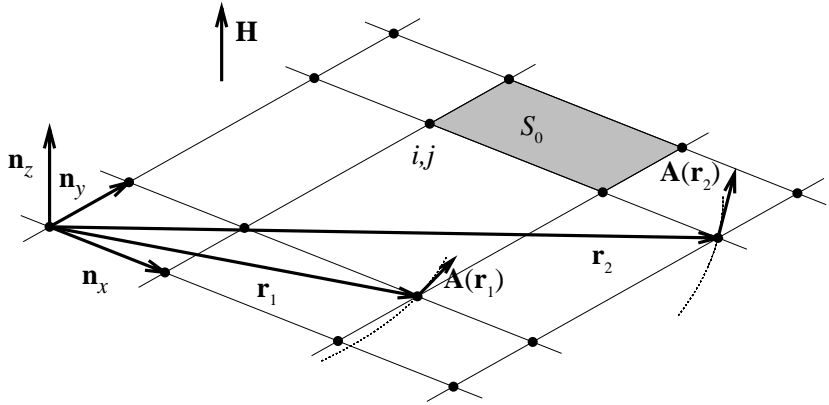}}
\caption[ ]
{\small Fermions in the magnetic field,
which appears within the mean--field treatment of the phase factors
in the Hamiltonian (\ref{038}).}
\label{fig5}
\end{figure}
and $L_x=L_y=L\to\infty$.
The corresponding magnetic field ${\bf{H}}({\bf{r}})$
immediately follows from Eq. (\ref{042})
\begin{eqnarray}
\label{043}
{\bf{H}}({\bf{r}})
={\mbox{rot}}{\bf{A}}({\bf{r}})
=\langle n_{{\bf{r}}^{\prime}}\rangle
\frac{\pi}{S_0}
\left\vert
\begin{array}{ccc}
{\bf{n}}_x & {\bf{n}}_y & {\bf{n}}_z \\
\frac{\partial}{\partial x} &
\frac{\partial}{\partial y} &
\frac{\partial}{\partial z} \\
-r_y & r_x & 0
\end{array}
\right\vert
=\langle n_{{\bf{r}}^{\prime}}\rangle
\frac{2\pi}{S_0}{\bf{n}}_z,
\end{eqnarray}
so that
${\bf{A}}({\bf{r}})=\frac{1}{2}
{\bf{H}}({\bf{r}})\times{\bf{r}}$.
The flux per elementary plaquette equals to
\begin{eqnarray}
\label{044}
\Phi_0={\bf{H}}({\bf{r}})\cdot S_0{\bf{n}}_z
=2\pi\langle n_{{\bf{r}}^{\prime}}\rangle=\pi.
\end{eqnarray}

The vector potential
${\bf{A}}({\bf{r}})$ (\ref{042}) shown schematically in Fig. 5
is not convenient to be concerned with.
Because of the gauge invariance
one may perform a gauge transformation
introducing a new vector potential
$\tilde{{\bf{A}}}({\bf{r}})$
which yields the same flux per elementary plaquette
$\Phi_0=\pi$.
Namely,
assume that
$\tilde{{\bf{A}}}({\bf{r}})$
is such that
\begin{eqnarray}
\label{045}
\alpha_{i+1,j}-\alpha_{i,j}
=\int_{i,j}^{i+1,j}{\mbox{d}}{\bf{r}}\cdot\tilde{{\bf{A}}}({\bf{r}})
=\pi,
\end{eqnarray}
whereas
\begin{eqnarray}
\label{046}
\alpha_{i+1,j+1}-\alpha_{i+1,j}
=\alpha_{i,j+1}-\alpha_{i+1,j+1}
=\alpha_{i,j}-\alpha_{i,j+1}
=0
\end{eqnarray}
(see Fig. 6).
From Eqs. (\ref{045}), (\ref{046})
one finds that
\begin{eqnarray}
\label{047}
\oint{\mbox{d}}{\bf{r}}\cdot\tilde{{\bf{A}}}({\bf{r}})
=\pi,
\end{eqnarray}
and on the other hand
\begin{eqnarray}
\label{048}
\oint{\mbox{d}}{\bf{r}}\cdot\tilde{{\bf{A}}}({\bf{r}})
=\int{\mbox{d}}{\bf{S}}\cdot{\mbox{rot}}\tilde{{\bf{A}}}({\bf{r}})
=\int{\mbox{d}}{\bf{S}}\cdot{\bf{H}}({\bf{r}})
=\Phi_0,
\end{eqnarray}
and hence the flux per elementary plaquette
$\Phi_0$ remains equal to $\pi$.

Let us turn back to the Hamiltonian (\ref{038}).
Only now we are in position to proceed 
with the statistical mechanics analysis. 
Within the frames of the introduced mean--field treatment
of the phase factors (\ref{045}), (\ref{046})
the Hamiltonian (\ref{038}) can be rewritten as
\begin{eqnarray}
\label{049}
H=
\sum_{\langle {\bf{i}},{\bf{j}} \rangle}
\frac{1}{2}J_{{\bf{i}},{\bf{j}}}
\left(d^+_{{\bf{i}}}d_{{\bf{j}}}
-d_{{\bf{i}}}d^+_{{\bf{j}}}\right)
+
H_Z+H_f,
\end{eqnarray}
where even for initially uniform lattice
$J=J^{\prime}=J_{\perp}=J_{\perp}^{\prime}$
in the Hamiltonian $H_{XY}$
one has
\begin{eqnarray}
\label{050}
J_{i,j;i+1,j}=-J,
\;\;\;
J_{i,j;i,j+1}=J,
\;\;\;
J_{i+1,j;i+2,j}=J,
\;\;\;
J_{i+1,j;i+1,j+1}=J
\end{eqnarray}
etc.
(see Fig. 6).
%
\begin{figure}
\epsfysize=100mm
\epsfclipon
\centerline{\epsffile{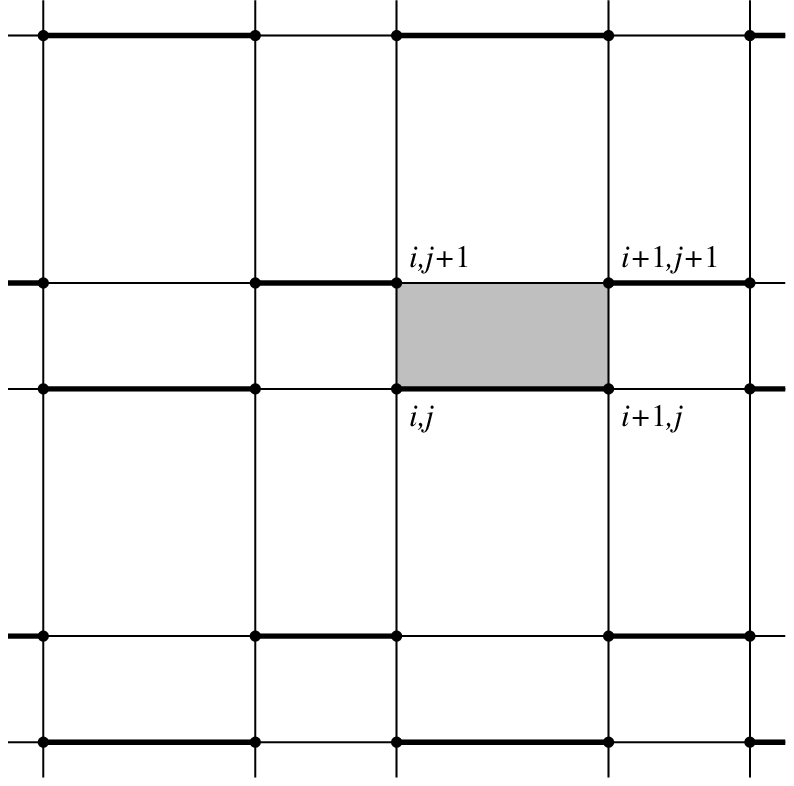}}
\caption[ ]
{\small Towards Eqs. (\ref{045}), (\ref{046})
and Eqs. (\ref{049}), (\ref{050}).}
\label{fig6}
\end{figure}
As can be seen from Eqs. (\ref{049}), (\ref{050})
the isotropic $XY$ model can be examined now
without making any other additional approximations
since it corresponds
to a model of tight--binding spinless fermions
on a bipartite square lattice
(see Section VI).
Conversely,
the Heisenberg model requires further approximations to proceed
because of the interaction between spinless fermions
(see Section VII).

\section{The Jordan--Wigner transformation in two dimensions
(E. Fradkin, 1989)}

The Fermi--Bose correspondence in two dimensions,
i.e., the Jordan--Wigner transformation
for two--dimensional spin--$\frac{1}{2}$ systems on a lattice,
was discussed even earlier \cite{004},
however,
without applications to the theory of concrete spin models.
Consider a system of spinless fermions,
i.e., a matter field,
$a({\bf{r}})$
on the sites of a square lattice (Fig. 7)
%
\begin{figure}
\epsfysize=80mm
\epsfclipon
\centerline{\epsffile{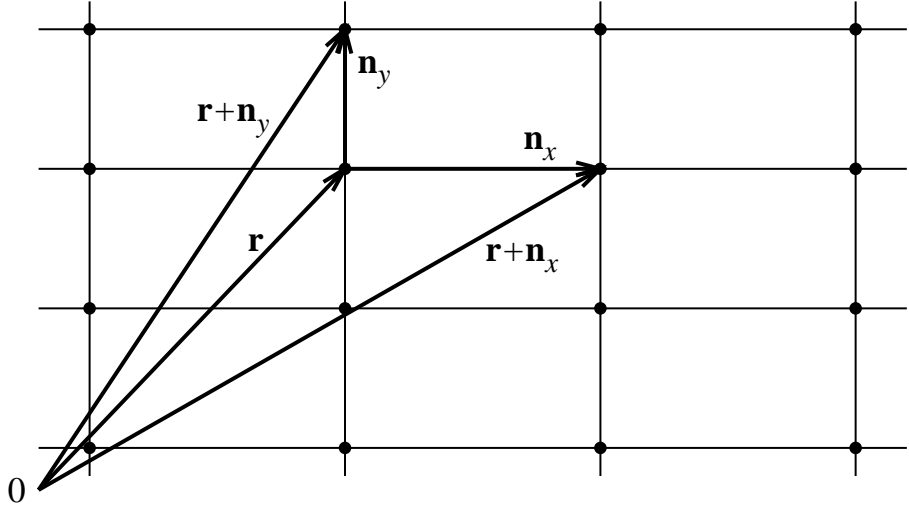}}
\caption[ ]
{\small Towards the Hamiltonian (\ref{051}).}
\label{fig7}
\end{figure}
and gauge field
$A_x({\bf{r}})$,
$A_y({\bf{r}})$
on the links of the lattice (Fig. 7).
The Hamiltonian of the system is
\begin{eqnarray}
\label{051}
H=t\sum_{{\bf{r}}}
\left(
a^+({\bf{r}})
{\mbox{e}}^{{\mbox{i}}A_x({\bf{r}})}
a({\bf{r}}+{\bf{n}}_x)
+a^+({\bf{r}})
{\mbox{e}}^{{\mbox{i}}A_y({\bf{r}})}
a({\bf{r}}+{\bf{n}}_y)
+{\mbox{h.c.}}
\right)
\end{eqnarray}
with the constraint
\begin{eqnarray}
\label{052}
a^+({\bf{r}})a({\bf{r}})
=\theta
\left(
A_y({\bf{r}}+{\bf{n}}_x)-A_y({\bf{r}})
-A_x({\bf{r}}+{\bf{n}}_y)+A_x({\bf{r}})
\right),
\end{eqnarray}
where $\theta$ is the parameter which will be defined later.
E. Fradkin showed \cite{004}
that the gauge field  can be eliminated
at the expense of a change in the commutation relations
of the matter field.
Namely,
introducing \cite{004}
the Jordan--Wigner operators
$\tilde{a}({\bf{r}})$, $\tilde{a}^+({\bf{r}})$
which obey
\begin{eqnarray}
\label{053}
\tilde{a}({\bf{r}}^{\prime})\tilde{a}^+({\bf{r}})
=\delta_{{\bf{r}}^{\prime},{\bf{r}}}
-{\mbox{e}}^{\frac{{\mbox{i}}}{2\theta}}
\tilde{a}^+({\bf{r}})\tilde{a}({\bf{r}}^{\prime})
\end{eqnarray}
with $\frac{1}{2\theta}=\pi$
the Hamiltonian (\ref{051}) is
\begin{eqnarray}
\label{054}
H=t\sum_{{\bf{r}}}
\left(
\tilde{a}^+({\bf{r}})\tilde{a}({\bf{r}}+{\bf{n}}_x)
+\tilde{a}^+({\bf{r}})\tilde{a}({\bf{r}}+{\bf{n}}_y)
+{\mbox{h.c.}}
\right).
\end{eqnarray}
Eq. (\ref{054}) is easily recognized  as the Hamiltonian
of the spin--$\frac{1}{2}$ isotropic $XY$ model
on a square lattice with exchange interaction
between nearest sites
$J=2t$
and the correspondence
$s^+_{\bf{r}}=\tilde{a}^+({\bf{r}})$,
$s^-_{\bf{r}}=\tilde{a}({\bf{r}})$,
$s^z_{\bf{r}}=\tilde{a}^+({\bf{r}})\tilde{a}({\bf{r}})
-\frac{1}{2}$.

Further discussions
on the extension of the Jordan--Wigner transformation
to three-- or more--dimensional cases
can be found in Ref. \onlinecite{015}.

\section{2D $s=\frac{1}{2}$ isotropic $XY$ model}

Let us show how the two--dimensional Jordan--Wigner transformation
with the mean--field treatment of the phase factors
can be used in the theory of spin models.
As a result we come to an approximate approach
to the study of 2D quantum spin models.

We begin with the 2D isotropic $XY$ model
($H_Z=H_f=0$)
considering for concreteness the case of
$J=J^{\prime}$,
$J_{\perp}=J_{\perp}^{\prime}$.
In accordance with (\ref{049}), (\ref{050})
we start from
\begin{eqnarray}
\label{055}
H_{XY}
=\frac{1}{2}\sum_{i=0}^{\infty}\sum_{j=0}^{\infty}
\left(J\left(-1\right)^{i+j}
\left(d^+_{i,j}d_{i+1,j}-d_{i,j}d^+_{i+1,j}\right)
+J_{\perp}
\left(d^+_{i,j}d_{i,j+1}-d_{i,j}d^+_{i,j+1}\right)
\right)
\nonumber\\
=\frac{1}{2}J
\left(\ldots
-a^+_{i,j}b_{i+1,j}+a_{i,j}b^+_{i+1,j}
+b^+_{i+1,j}a_{i+2,j}-b_{i+1,j}a^+_{i+2,j}
+\ldots\right)
\nonumber\\
+\frac{1}{2}J_{\perp}
\left(\ldots
+a^+_{i,j}b_{i,j+1}-a_{i,j}b^+_{i,j+1}
+b^+_{i+1,j}a_{i+1,j+1}-b_{i+1,j}a^+_{i+1,j+1}
+\ldots\right),
\end{eqnarray}
where, 
to emphasize a bipartite character 
of the square lattice that appeared,
we have introduced notations 
$a_{i,j}=d_{i,j}$,
$b_{i+1,j}=d_{i+1,j}$
etc..

Notwithstanding the fact that we have an approximate theory, 
it contains the exact result in 1D limit  
if one puts in Eq. (\ref{055})
either $J_{\perp}=0$ or $J=0$
coming to a system of noninteracting chains 
extended in either horizontal or vertical direction,
respectively.  
In the latter case Eq. (\ref{055}) 
corresponds to a system of noninteracting chains 
each with the $XY$ part of the Hamiltonian (\ref{013})
\begin{eqnarray}
\label{055aa}
H_{XY}(i)=\sum_{j=0}^{\infty}
\frac{1}{2}J_{\perp}
\left(d^+_{i,j}d_{i,j+1}-d_{i,j}d^+_{i,j+1}\right).
\end{eqnarray}
In the former case one gets a system of noninteracting chains 
each with the Hamiltonian
\begin{eqnarray}
\label{055a}
H_{XY}(j)=\left(-1\right)^j\sum_{i=0}^{\infty}
\frac{1}{2}
J\left(-1\right)^i
\left(d^+_{i,j}d_{i+1,j}-d_{i,j}d^+_{i+1,j}\right)
\end{eqnarray}
and to recover the 1D limit explicitly 
the transformation
\begin{eqnarray}
\label{055b}
d^+_{i,j}={\mbox{e}}^{{\mbox{i}}\pi\psi_i}f^+_i,
d_{i+1,j}={\mbox{e}}^{-{\mbox{i}}\pi\psi_{i+1}}f_{i+1},
\ldots,
\;\;\;
\psi_0=0, 
\psi_{i+1}=\psi_i+i
\end{eqnarray}
(e.g., $d^+_{0,j}=f^+_0$, 
$d^+_{1,j}=-f^+_1$,
$d^+_{2,j}=-f^+_2$,
$d^+_{3,j}=f^+_3$
etc.)
should be performed.

After the Fourier transformation
\begin{eqnarray}
\label{056}
d_{i,j}=\frac{1}{\sqrt{N_xN_y}}
\sum_{k_x,k_y}
{\mbox{e}}^{{\mbox{i}}
\left(k_xi+k_yj\right)}
d_{k_x,k_y}
\end{eqnarray}
etc. or in short
\begin{eqnarray}
\label{057}
d_{{\bf{i}}}=\frac{1}{\sqrt{N}}
\sum_{{\bf{k}}}
{\mbox{e}}^{{\mbox{i}}
{\bf{k}}\cdot{\bf{i}}}
d_{{\bf{k}}},
\;\;\;
d^+_{{\bf{i}}}=\frac{1}{\sqrt{N}}
\sum_{{\bf{k}}}
{\mbox{e}}^{-{\mbox{i}}
{\bf{k}}\cdot{\bf{i}}}
d^+_{{\bf{k}}},
\;\;\;
d_{{\bf{k}}}=\frac{1}{\sqrt{N}}
\sum_{{\bf{i}}}
{\mbox{e}}^{-{\mbox{i}}
{\bf{k}}\cdot{\bf{i}}}
d_{{\bf{i}}},
\;\;\;
d^+_{{\bf{k}}}=\frac{1}{\sqrt{N}}
\sum_{{\bf{i}}}
{\mbox{e}}^{{\mbox{i}}
{\bf{k}}\cdot{\bf{i}}}
d^+_{{\bf{i}}},
\nonumber\\
k_x=\frac{2\pi}{N_x}n_x,
\;\;\;
n_x=-\frac{N_x}{2},-\frac{N_x}{2}+1,\ldots,\frac{N_x}{2}-1,
\;\;\;
k_y=\frac{2\pi}{N_y}n_y,
\;\;\;
n_y=-\frac{N_y}{2},-\frac{N_y}{2}+1,\ldots,\frac{N_y}{2}-1,
\nonumber\\
\left\{d_{{\bf{k}}_1},d^+_{{\bf{k}}_2}\right\}
=\delta_{{\bf{k}}_1,{\bf{k}}_2},
\;\;\;
\left\{d_{{\bf{k}}_1},d_{{\bf{k}}_2}\right\}
=\left\{d^+_{{\bf{k}}_1},d^+_{{\bf{k}}_2}\right\}
=0
\end{eqnarray}
($N_x$, $N_y$ are even),
the Hamiltonian (\ref{055}) becomes
\begin{eqnarray}
\label{058}
H_{XY}
=\frac{1}{2}
\sum_{{\bf{k}}}
\left(
{\mbox{i}}J\sin k_x
\left(
b^+_{{\bf{k}}}a_{{\bf{k}}}
-a^+_{{\bf{k}}}b_{{\bf{k}}}
\right)
+
J_{\perp}\cos k_y
\left(
b^+_{{\bf{k}}}b_{{\bf{k}}}
-a^+_{{\bf{k}}}a_{{\bf{k}}}
\right)
\right)
\nonumber\\
=\frac{1}{2}
\sum_{{\bf{k}}}
\vert E_{{\bf{k}}}\vert
\left(
\cos\gamma_{{\bf{k}}}
\left(
b^+_{{\bf{k}}}b_{{\bf{k}}}
-a^+_{{\bf{k}}}a_{{\bf{k}}}
\right)
+
{\mbox{i}}\sin\gamma_{{\bf{k}}}
\left(
b^+_{{\bf{k}}}a_{{\bf{k}}}
-a^+_{{\bf{k}}}b_{{\bf{k}}}
\right)
\right)
\end{eqnarray}
where we have used the notations
\begin{eqnarray}
\label{059}
b^+_{{\bf{k}}}=d^+_{k_x,k_y},
\;\;\;
a_{{\bf{k}}}=d_{k_x\pm\pi,k_y\pm\pi},
\;\;\;
\ldots,
\nonumber\\
E_{{\bf{k}}}=J_{\perp}\cos k_y
+{\mbox{i}}J\sin k_x
=\vert E_{{\bf{k}}}\vert
{\mbox{e}}^{{\mbox{i}}\gamma_{{\bf{k}}}},
\nonumber\\
\vert E_{{\bf{k}}}\vert
=\sqrt{J_{\perp}^2\cos^2k_y
+J^2\sin^2k_x},
\;\;\;
\cos\gamma_{{\bf{k}}}
=\frac{J_{\perp}\cos k_y}{\vert E_{{\bf{k}}}\vert}
\;\;\;
\sin\gamma_{{\bf{k}}}
=\frac{J\sin k_x}{\vert E_{{\bf{k}}}\vert}.
\end{eqnarray}
The Hamiltonian (\ref{058}) can be rewritten as
\begin{eqnarray}
\label{059a}
H_{XY}
=\sum_{{\bf{k}}}{^{\prime}}
\vert E_{{\bf{k}}}\vert
\left(
\cos\gamma_{{\bf{k}}}
\left(
b^+_{{\bf{k}}}b_{{\bf{k}}}
-a^+_{{\bf{k}}}a_{{\bf{k}}}
\right)
+
{\mbox{i}}\sin\gamma_{{\bf{k}}}
\left(
b^+_{{\bf{k}}}a_{{\bf{k}}}
-a^+_{{\bf{k}}}b_{{\bf{k}}}
\right)
\right)
\end{eqnarray}
where the prime denotes that ${\bf{k}}$ 
varies in the region shown in Fig. 8b.
%
\begin{figure}
\epsfysize=80mm
\epsfclipon
\centerline{\epsffile{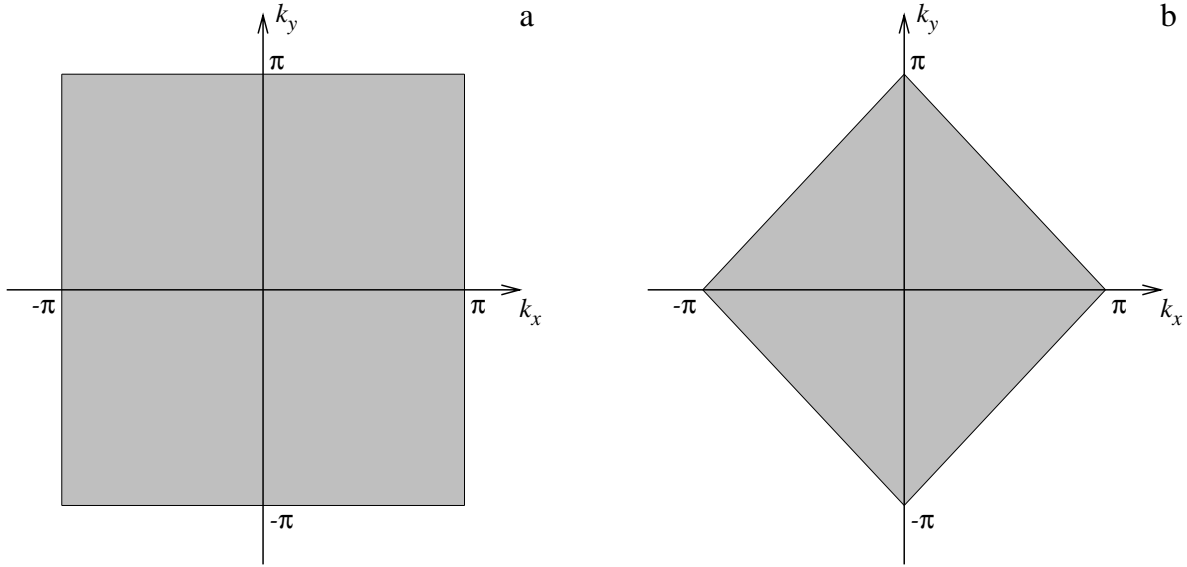}}
\caption[ ]
{\small The region in which ${\bf{k}}$ varies 
in the sum in Eq. (\ref{058}) (see Eq. (\ref{057})) (a)
and in Eq. (\ref{059a}) (b).}
\label{fig8}
\end{figure}

At last one introduces the operators
\begin{eqnarray}
\label{060}
\alpha_{{\bf{k}}}
=\cos\frac{\gamma_{{\bf{k}}}}{2}b_{{\bf{k}}}
+{\mbox{i}}\sin\frac{\gamma_{{\bf{k}}}}{2}a_{{\bf{k}}},
\;\;\;
\beta_{{\bf{k}}}
=\sin\frac{\gamma_{{\bf{k}}}}{2}b_{{\bf{k}}}
-{\mbox{i}}\cos\frac{\gamma_{{\bf{k}}}}{2}a_{{\bf{k}}},
\nonumber\\
\left\{\alpha_{{\bf{k}}_1},\alpha^+_{{\bf{k}}_2}\right\}
=\delta_{{\bf{k}}_1,{\bf{k}}_2},
\;\;\;
\left\{\beta_{{\bf{k}}_1},\beta^+_{{\bf{k}}_2}\right\}
=\delta_{{\bf{k}}_1,{\bf{k}}_2},
\;\;\;
\left\{\alpha_{{\bf{k}}_1},\beta^+_{{\bf{k}}_2}\right\}
=\left\{\beta_{{\bf{k}}_1},\alpha^+_{{\bf{k}}_2}\right\}
=0,
\end{eqnarray}
etc.
to get from Eq. (\ref{059a}) the final form
of the 2D spin--$\frac{1}{2}$ isotropic $XY$ model Hamiltonian
in fermionic language
within approximation (\ref{041})
\begin{eqnarray}
\label{061}
H_{XY}
=\sum_{{\bf{k}}}{^{\prime}}
\Lambda_{{\bf{k}}}
\left(
\alpha^+_{{\bf{k}}}\alpha_{{\bf{k}}}
-\beta^+_{{\bf{k}}}\beta_{{\bf{k}}}
\right),
\nonumber\\
\Lambda_{{\bf{k}}}=\vert E_{{\bf{k}}}\vert
=\sqrt{J^2\sin^2k_x+
J_{\perp}^2\cos^2k_y}\ge 0.
\end{eqnarray}

It is easy now to calculate
the thermodynamic functions of the spin model
which correspond to Eq. (\ref{061}).
For example, the ground state energy per site is
\begin{eqnarray}
\label{062}
\frac{E_0}{N}
=-\int_{-\pi}^{\pi}\frac{{\mbox{d}}k_x}{2\pi}
\int_{-\pi-\vert k_x\vert}^{\pi-\vert k_x\vert}
\frac{{\mbox{d}}k_y}{2\pi}
\sqrt{J^2\sin^2k_x+
J_{\perp}^2\cos^2k_y}
\nonumber\\
=-\frac{1}{2}
\int_{-\pi}^{\pi}\frac{{\mbox{d}}k_x}{2\pi}
\int_{-\pi}^{\pi}\frac{{\mbox{d}}k_y}{2\pi}
\sqrt{J^2\sin^2k_x+
J_{\perp}^2\cos^2k_y}.
\end{eqnarray}
In 1D limit 
($J_{\perp}=0$ or $J=0$)
Eq. (\ref{062})
becomes
\begin{eqnarray}
\label{062a}
\frac{E_0}{N}
=-\frac{1}{2}
\int_{-\pi}^{\pi}\frac{{\mbox{d}}k_x}{2\pi}
\vert J\sin k_x\vert
=-\frac{1}{2}
\int_{-\pi}^{\pi}\frac{{\mbox{d}}k_y}{2\pi}
\vert J_{\perp}\cos k_y\vert
=-\frac{\vert J\vert}{\pi}
\end{eqnarray}
that is the well known exact result.

It is remarkable to note
that the Hamiltonian (\ref{061})
(and hence the ground state energy per site (\ref{062}))
arises also
for the 2D spin--$\frac{1}{2}$ Heisenberg model
and is known as
the {\it {uniform flux}}
($\langle n_{{\bf{r}}}\rangle=\frac{1}{2}$)
solution \cite{005}.

\section{2D $s=\frac{1}{2}$ Heisenberg model}

In the remainder of the paper
we consider
the 2D spin--$\frac{1}{2}$ isotropic Heisenberg model
(\ref{001}), (\ref{002})
treating the phase factors
which appear after making use of the 2D Jordan--Wigner transformation
within the frames of the mean--field approximation (\ref{041}).
The Heisenberg model besides
the $H_{XY}$ term (\ref{055})
includes the $H_Z$ term
which contains
\begin{eqnarray}
\label{063}
d_{\bf{i}}^+d_{\bf{i}}d_{\bf{j}}^+d_{\bf{j}}
-\frac{1}{2}d_{\bf{i}}^+d_{\bf{i}}
-\frac{1}{2}d_{\bf{j}}^+d_{\bf{j}}
+\frac{1}{4}.
\end{eqnarray}
Thus,
the Jordan--Wigner spinless fermions interact
and further approximations are required.
The first term in (\ref{063}) can be changed by
\begin{eqnarray}
\label{064}
d_{\bf{i}}^+d_{\bf{i}}d_{\bf{j}}^+d_{\bf{j}}
\rightarrow
d_{\bf{i}}^+d_{\bf{i}}\langle d_{\bf{j}}^+d_{\bf{j}}\rangle
+\langle d_{\bf{i}}^+d_{\bf{i}}\rangle d_{\bf{j}}^+d_{\bf{j}}
-\langle d_{\bf{i}}^+d_{\bf{i}}\rangle
\langle d_{\bf{j}}^+d_{\bf{j}}\rangle
\nonumber\\
+d_{\bf{i}}^+d_{\bf{j}}\langle d_{\bf{i}}d_{\bf{j}}^+\rangle
+\langle d_{\bf{i}}^+d_{\bf{j}}\rangle d_{\bf{i}}d_{\bf{j}}^+
-\langle d_{\bf{i}}^+d_{\bf{j}}\rangle
\langle d_{\bf{i}}d_{\bf{j}}^+\rangle
\nonumber\\
=d_{\bf{i}}^+d_{\bf{i}}\langle n_{\bf{j}} \rangle
+d_{\bf{j}}^+d_{\bf{j}}\langle n_{\bf{i}} \rangle
-\langle n_{\bf{i}} \rangle
\langle n_{\bf{j}} \rangle
\nonumber\\
+d_{\bf{i}}^+d_{\bf{j}} \Delta_{{\bf{i}},{\bf{j}}}
{\mbox{e}}^{{\mbox{i}}\theta_{{\bf{i}},{\bf{j}}}}
-d_{\bf{i}}d_{\bf{j}}^+ \Delta_{{\bf{j}},{\bf{i}}}
{\mbox{e}}^{{\mbox{i}}\theta_{{\bf{j}},{\bf{i}}}}
+\Delta_{{\bf{i}},{\bf{j}}}\Delta_{{\bf{j}},{\bf{i}}}
{\mbox{e}}^{{\mbox{i}}
\left(\theta_{{\bf{i}},{\bf{j}}}+\theta_{{\bf{j}},{\bf{i}}}\right)},
\end{eqnarray}
where we have introduced the notation
\begin{eqnarray}
\label{065}
\langle d_{\bf{i}}d_{\bf{j}}^+\rangle
=\Delta_{{\bf{i}},{\bf{j}}}
{\mbox{e}}^{{\mbox{i}}\theta_{{\bf{i}},{\bf{j}}}}
=-\langle d_{\bf{j}}^+d_{\bf{i}}\rangle.
\end{eqnarray}
In accordance with (\ref{064})
there may be four ways to treat the Ising interaction,
i.e. assuming either
\begin{eqnarray}
\label{066}
\Delta_{{\bf{i}},{\bf{j}}}=0,
\;\;\;
m=\langle s_{\bf{j}}^z\rangle
=\langle n_{\bf{j}} \rangle
-\frac{1}{2}=0
\end{eqnarray}
or
\begin{eqnarray}
\label{067}
\Delta_{{\bf{i}},{\bf{j}}}\ne 0,
\;\;\;
m= 0
\end{eqnarray}
or
\begin{eqnarray}
\label{068}
\Delta_{{\bf{i}},{\bf{j}}}=0,
\;\;\;
m\ne 0
\end{eqnarray}
or
\begin{eqnarray}
\label{069}
\Delta_{{\bf{i}},{\bf{j}}}\ne 0,
\;\;\;
m\ne 0.
\end{eqnarray}

The first possibility (\ref{066})
yielding the {\it {uniform flux}} solution
was considered in the previous Section
(Eqs. (\ref{061}), (\ref{062})).

Let us consider the second possibility (\ref{067}).
In such a case the Ising term becomes
\begin{eqnarray}
\label{070}
H_Z=
\sum_{\langle {\bf{i}},{\bf{j}}\rangle}
J_{{\bf{i}},{\bf{j}}}
\left(
\Delta_{{\bf{i}},{\bf{j}}}
{\mbox{e}}^{{\mbox{i}}\theta_{{\bf{i}},{\bf{j}}}}
d_{{\bf{i}}}^+d_{{\bf{j}}}
-
\Delta_{{\bf{j}},{\bf{i}}}
{\mbox{e}}^{{\mbox{i}}\theta_{{\bf{j}},{\bf{i}}}}
d_{{\bf{i}}}d_{{\bf{j}}}^+
+\Delta_{{\bf{i}},{\bf{j}}}\Delta_{{\bf{j}},{\bf{i}}}
{\mbox{e}}^{{\mbox{i}}
\left(\theta_{{\bf{i}},{\bf{j}}}
+\theta_{{\bf{j}},{\bf{i}}}\right)}
\right),
\end{eqnarray}
where
to get the {\it {in--phase flux}} solution
one puts
\begin{eqnarray}
\label{071}
\Delta_{i,j;i+1,j}
=\Delta_{i+1,j;i,j}
=\Delta_{i+1,j;i+2,j}
=\Delta_{i+2,j;i+1,j}
=Q,
\nonumber\\
{\mbox{e}}^{{\mbox{i}}\theta_{i,j;i+1,j}}
={\mbox{e}}^{{\mbox{i}}\theta_{i+1,j;i,j}}=-1,
\;\;\;
{\mbox{e}}^{{\mbox{i}}\theta_{i+1,j;i+2,j}}
={\mbox{e}}^{{\mbox{i}}\theta_{i+2,j;i+1,j}}=1,
\nonumber\\
\Delta_{i,j;i,j+1}
=\Delta_{i,j+1;i,j}
=\Delta_{i+1,j;i+1,j+1}
=\Delta_{i+1,j+1;i+1,j}
=P,
\nonumber\\
{\mbox{e}}^{{\mbox{i}}\theta_{i,j;i,j+1}}
={\mbox{e}}^{{\mbox{i}}\theta_{i,j+1;i,j}}
={\mbox{e}}^{{\mbox{i}}\theta_{i+1,j;i+1,j+1}}
={\mbox{e}}^{{\mbox{i}}\theta_{i+1,j+1;i+1,j}}=1
\end{eqnarray}
(see Fig. 9)
%
\begin{figure}
\epsfysize=80mm
\epsfclipon
\centerline{\epsffile{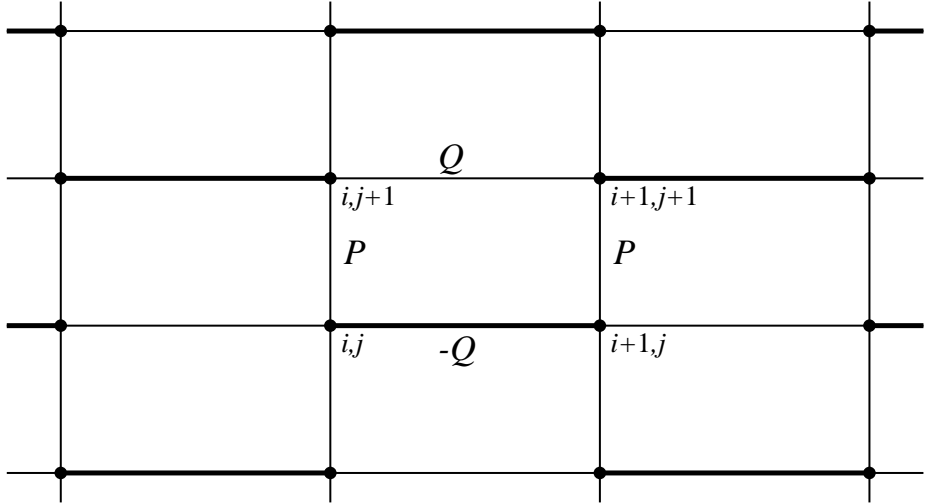}}
\caption[ ]
{\small Towards the in--phase flux solution
for the Heisenberg model;
the values of
$\Delta_{{\bf{i}},{\bf{j}}}
{\mbox{e}}^{{\mbox{i}}\theta_{{\bf{i}},{\bf{j}}}}$
are attached to the bonds.}
\label{fig9}
\end{figure}
and the parameters $Q$ and $P$ are calculated self--consistently
(see Eqs. (\ref{075}), (\ref{076}) below).
Now the Heisenberg model Hamiltonian
only slightly differs from that of the $XY$ model (\ref{055})
becoming
\begin{eqnarray}
\label{072}
H=
\frac{1}{2}J\left(1+2Q\right)
\left(\ldots
-a^+_{i,j}b_{i+1,j}+a_{i,j}b^+_{i+1,j}
+b^+_{i+1,j}a_{i+2,j}-b_{i+1,j}a^+_{i+2,j}
+\ldots\right)
\nonumber\\
+\frac{1}{2}J_{\perp}\left(1+2P\right)
\left(\ldots
+a^+_{i,j}b_{i,j+1}-a_{i,j}b^+_{i,j+1}
+b^+_{i+1,j}a_{i+1,j+1}-b_{i+1,j}a^+_{i+1,j+1}
+\ldots\right)
\nonumber\\
+NJQ^2+NJ_{\perp}P^2.
\end{eqnarray}
Acting along the line described in the Section VI
one finds that
\begin{eqnarray}
\label{073}
H=\sum_{{\bf{k}}}{^{\prime}}
\Lambda_{{\bf{k}}}
\left(\alpha^+_{{\bf{k}}}\alpha_{{\bf{k}}}
-\beta^+_{{\bf{k}}}\beta_{{\bf{k}}}\right)
+NJQ^2+NJ_{\perp}P^2,
\nonumber\\
\Lambda_{{\bf{k}}}
=\sqrt{J^2\left(1+2Q\right)^2\sin^2k_x+
J_{\perp}^2\left(1+2P\right)^2\cos^2k_y}\ge 0.
\end{eqnarray}
The ground state energy per site is given by
\begin{eqnarray}
\label{074}
\frac{E_0}{N}
=-\frac{1}{2}
\int_{-\pi}^{\pi}\frac{{\mbox{d}}k_x}{2\pi}
\int_{-\pi}^{\pi}\frac{{\mbox{d}}k_y}{2\pi}
\sqrt{J^2\left(1+2Q\right)^2\sin^2k_x+
J_{\perp}^2\left(1+2P\right)^2\cos^2k_y}
\nonumber\\
+JQ^2+J_{\perp}P^2,
\end{eqnarray}
where the parameters $Q$ and $P$
are determined from the conditions
$\frac{\partial}{\partial Q}\frac{E_0}{N}=0$
and
$\frac{\partial}{\partial P}\frac{E_0}{N}=0$,
i.e.,
\begin{eqnarray}
\label{075}
Q=\frac{1}{2}
\int_{-\pi}^{\pi}\frac{{\mbox{d}}k_x}{2\pi}
\int_{-\pi}^{\pi}\frac{{\mbox{d}}k_y}{2\pi}
\frac{J\sin^2 k_x\left(1+2Q\right)}
{\sqrt{J^2\left(1+2Q\right)^2\sin^2k_x+
J_{\perp}^2\left(1+2P\right)^2\cos^2k_y}}
\end{eqnarray}
and
\begin{eqnarray}
\label{076}
P=\frac{1}{2}
\int_{-\pi}^{\pi}\frac{{\mbox{d}}k_x}{2\pi}
\int_{-\pi}^{\pi}\frac{{\mbox{d}}k_y}{2\pi}
\frac{J_{\perp}\cos^2 k_y\left(1+2P\right)}
{\sqrt{J^2\left(1+2Q\right)^2\sin^2k_x+
J_{\perp}^2\left(1+2P\right)^2\cos^2k_y}}.
\end{eqnarray}

Treating the Ising term (\ref{064})
within assumptions (\ref{068}) or (\ref{069})
one assumes the N\'{e}el order
with the sublattice magnetizations $m$ and $-m$
\begin{eqnarray}
\label{077}
\langle n_{i,j}\rangle=\langle n_{i+1,j+1}\rangle=\ldots=
m+\frac{1}{2},
\;\;\;
\langle n_{i,j+1}\rangle=\langle n_{i+1,j}\rangle=\ldots=
-m+\frac{1}{2}
\end{eqnarray}
(see Fig. 10).
%
\begin{figure}
\epsfysize=80mm
\epsfclipon
\centerline{\epsffile{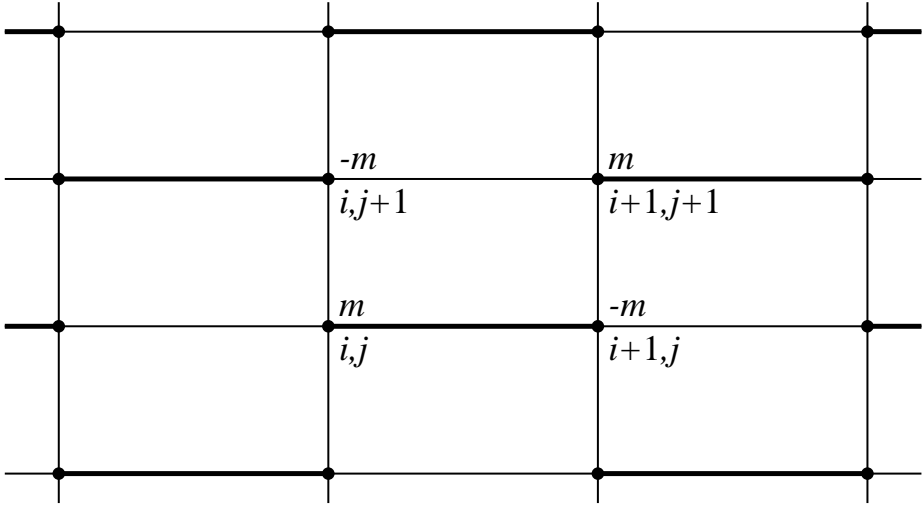}}
\caption[ ]
{\small The N\'{e}el order
for the Heisenberg model;
the values of magnetization
$\langle s^z_{{\bf{i}}} \rangle$
are attached to the sites.}
\label{fig10}
\end{figure}
(The given assumption, by the way,
apparently contradicts the mean--field approximation (\ref{041})
for the phase factors;
this inconsistency, however,
to our best knowledge
has not yet been discussed.)
Therefore,
case (\ref{068}),
i.e., the uniform flux with the N\'{e}el order,
is called the {\it {N\'{e}el flux}} solution
and case (\ref{069}),
i.e., the in--flux with the N\'{e}el order,
is called the {\it {in--phase N\'{e}el flux}} solution.
In the latter case the Hamiltonian reads
\begin{eqnarray}
\label{078}
H=
\frac{1}{2}J\left(1+2Q\right)
\left(\ldots -a^+_{i,j}b_{i+1,j}+a_{i,j}b^+_{i+1,j}
+b_{i+1,j}^+a_{i+2,j}-b_{i+1,j}a^+_{i+2,j}
+\ldots\right)
\nonumber\\
+
\frac{1}{2}J_{\perp}\left(1+2P\right)
\left(\ldots +a^+_{i,j}b_{i,j+1}-a_{i,j}b^+_{i,j+1}
+b_{i+1,j}^+a_{i+1,j+1}-b_{i+1,j}a^+_{i+1,j+1}+\ldots\right)
\nonumber\\
+
J\left(
\ldots -ma^+_{i,j}a_{i,j}+mb^+_{i+1,j}b_{i+1,j}
+mb^+_{i+1,j}b_{i+1,j}-ma^+_{i+2,j}a_{i+2,j}+\ldots
\right)
\nonumber\\
+
J_{\perp}\left(
\ldots -ma^+_{i,j}a_{i,j}+mb^+_{i,j+1}b_{i,j+1}
+mb^+_{i+1,j}b_{i+1,j}-ma^+_{i+1,j+1}a_{i+1,j+1}+\ldots
\right)
\nonumber\\
+NJQ^2
+NJ_{\perp}P^2
+N\left(J+J_{\perp}\right)m^2.
\end{eqnarray}
Performing the Fourier transformation (\ref{057})
one gets
\begin{eqnarray}
\label{079}
H=\frac{1}{2}
\sum_{{\bf{k}}}
\left(
\left( 2\left( J+J_{\perp}\right)m
+{\mbox{i}}J\left( 1+2Q\right)\sin k_x\right)
b^+_{{\bf{k}}}a_{{\bf{k}}}
\right.
\nonumber\\
\left.
+
\left( 2\left( J+J_{\perp}\right)m
-{\mbox{i}}J\left( 1+2Q\right)\sin k_x\right)
a^+_{{\bf{k}}}b_{{\bf{k}}}
\right.
\nonumber\\
\left.
+
J_{\perp}\left(1+2P\right)\cos k_y
\left(
b^+_{{\bf{k}}}b_{{\bf{k}}}
-a^+_{{\bf{k}}}a_{{\bf{k}}}
\right)
\right)
\nonumber\\
+NJQ^2+NJ_{\perp}P^2+N\left(J+J_{\perp}\right)m^2
\nonumber\\
=\sum_{{\bf{k}}}{^{\prime}}
\left(
\left( 2\left( J+J_{\perp}\right)m
+{\mbox{i}}J\left( 1+2Q\right)\sin k_x\right)
b^+_{{\bf{k}}}a_{{\bf{k}}}
\right.
\nonumber\\
\left.
+
\left( 2\left( J+J_{\perp}\right)m
-{\mbox{i}}J\left( 1+2Q\right)\sin k_x\right)
a^+_{{\bf{k}}}b_{{\bf{k}}}
\right.
\nonumber\\
\left.
+
J_{\perp}\left(1+2P\right)\cos k_y
\left(
b^+_{{\bf{k}}}b_{{\bf{k}}}
-a^+_{{\bf{k}}}a_{{\bf{k}}}
\right)
\right)
\nonumber\\
+NJQ^2+NJ_{\perp}P^2+N\left(J+J_{\perp}\right)m^2,
\end{eqnarray}
where
$b^+_{{\bf{k}}}=d^+_{k_x,k_y}$,
$a_{{\bf{k}}}=d_{k_x\pm\pi,k_y\pm\pi}$,
etc..
Introducing the operators
\begin{eqnarray}
\label{080}
\tilde{a}_{{\bf{k}}}=a_{{\bf{k}}}
{\mbox{e}}^{{\mbox{i}}\delta_{{\bf{k}}}},
\;\;\;
\tilde{a}^+_{{\bf{k}}}=a^+_{{\bf{k}}}
{\mbox{e}}^{-{\mbox{i}}\delta_{{\bf{k}}}},
\;\;\;
\tilde{b}_{{\bf{k}}}=b_{{\bf{k}}},
\;\;\;
\tilde{b}^+_{{\bf{k}}}=b^+_{{\bf{k}}};
\nonumber\\
2\left(J+J_{\perp}\right)m
\pm{\mbox{i}}J\left(1+2Q\right)\sin k_x=
\sqrt{4\left(J+J_{\perp}\right)^2m^2
+J^2\left(1+2Q\right)^2\sin^2 k_x}
\;
{\mbox{e}}^{\pm{\mbox{i}}\delta_{{\bf{k}}}}
\end{eqnarray}
and then the operators
\begin{eqnarray}
\label{081}
\alpha_{{\bf{k}}}
=\cos\frac{\omega_{{\bf{k}}}}{2}\tilde{b}_{{\bf{k}}}
+\sin\frac{\omega_{{\bf{k}}}}{2}\tilde{a}_{{\bf{k}}},
\;\;\;
\beta_{{\bf{k}}}
=\sin\frac{\omega_{{\bf{k}}}}{2}\tilde{b}_{{\bf{k}}}
-\cos\frac{\omega_{{\bf{k}}}}{2}\tilde{a}_{{\bf{k}}};
\nonumber\\
\cos \omega_{{\bf{k}}}
=\frac{J_{\perp}\left(1+2P\right)\cos k_y}
{\vert E_{{\bf{k}}}\vert},
\;\;\;
\sin \omega_{{\bf{k}}}
=\frac{\sqrt{4\left(J+J_{\perp}\right)^2m^2
+J^2\left(1+2Q\right)^2\sin^2 k_x}}
{\vert E_{{\bf{k}}}\vert},
\nonumber\\
\vert E_{{\bf{k}}}\vert
=\sqrt{4\left(J+J_{\perp}\right)^2m^2
+J^2\left(1+2Q\right)^2\sin^2 k_x
+J_{\perp}^2\left(1+2P\right)^2\cos^2 k_y}
\end{eqnarray}
one gets the final form of the Heisenberg Hamiltonian
in fermionic language
\begin{eqnarray}
\label{082}
H=\sum_{{\bf{k}}}{^{\prime}}
\Lambda_{{\bf{k}}}
\left(
\alpha^+_{{\bf{k}}}\alpha_{{\bf{k}}}
-\beta^+_{{\bf{k}}}\beta_{{\bf{k}}}
\right)
+NJQ^2+NJ_{\perp}P^2+N\left(J+J_{\perp}\right)m^2
\end{eqnarray}
with $\Lambda_{{\bf{k}}}=\vert E_{{\bf{k}}}\vert\ge 0$
defined by Eq. (\ref{081}).

The ground state energy per site
which follows from (\ref{082}) reads
\begin{eqnarray}
\label{083}
\frac{E_0}{N}
=-\frac{1}{2}
\int_{-\pi}^{\pi}\frac{{\mbox{d}}k_x}{2\pi}
\int_{-\pi}^{\pi}\frac{{\mbox{d}}k_y}{2\pi}
\sqrt{4\left(J+J_{\perp}\right)^2m^2
+J^2\left(1+2Q\right)^2\sin^2k_x+
J_{\perp}^2\left(1+2P\right)^2\cos^2k_y}
\nonumber\\
+JQ^2+J_{\perp}P^2+\left(J+J_{\perp}\right)m^2,
\end{eqnarray}
where  the values of the introduced parameters
are determined
by minimizing $\frac{E_0}{N}$ (\ref{083})
with respect to $Q$, $P$ and $m$
\begin{eqnarray}
\label{084}
Q=\frac{1}{2}
\int\frac{{\mbox{d}}{\bf{k}}}
{\left(2\pi\right)^2}
\frac{J\sin^2 k_x\left(1+2Q\right)}
{\sqrt{4\left(J+J_{\perp}\right)^2m^2
+J^2\left(1+2Q\right)^2\sin^2k_x+
J_{\perp}^2\left(1+2P\right)^2\cos^2k_y}},
\end{eqnarray}
\begin{eqnarray}
\label{085}
P=\frac{1}{2}
\int\frac{{\mbox{d}}{\bf{k}}}
{\left(2\pi\right)^2}
\frac{J_{\perp}\cos^2 k_y\left(1+2P\right)}
{\sqrt{4\left(J+J_{\perp}\right)^2m^2
+J^2\left(1+2Q\right)^2\sin^2k_x+
J_{\perp}^2\left(1+2P\right)^2\cos^2k_y}},
\end{eqnarray}
\begin{eqnarray}
\label{086}
2m=
\int\frac{{\mbox{d}}{\bf{k}}}
{\left(2\pi\right)^2}
\frac{2\left(J+J_{\perp}\right)m}
{\sqrt{4\left(J+J_{\perp}\right)^2m^2
+J^2\left(1+2Q\right)^2\sin^2k_x+
J_{\perp}^2\left(1+2P\right)^2\cos^2k_y}}.
\end{eqnarray}

The N\'{e}el flux solution (see Eq. (\ref{068})) is given
by Eqs. (\ref{078}), (\ref{082}), (\ref{081}),
(\ref{083}), (\ref{086})
in which $P$ and $Q$ are equal to zero.

\section{The 2D $s=\frac{1}{2}$ Heisenberg model:
A comparison of some results}

In the final Section we want to list out some problems
of the two--dimensional spin models theory
which were attacked
with the exploiting of the 2D Jordan--Wigner transformation.

The 2D antiferromagnetic Heisenberg model
with $J=J^{\prime}=J_{\perp}=J_{\perp}^{\prime}$
(without external field)
was considered in Refs. \onlinecite{005,007,006}.
The main results obtained concern
the ground state energy \cite{005,007},
the specific heat \cite{007}
and the Raman spectrum \cite{006,007}.
A comparison with some experimental data for La$_2$CuO$_4$
was given.

In Ref. \onlinecite{008}
the effects of the interchain interaction
on the one--dimensional
spin--$\frac{1}{2}$ antiferromagnetic Heisenberg model
were examined.
For this purpose the Heisenberg model
with $J=J^{\prime}$, $J_{\perp}=J_{\perp}^{\prime}$
was considered
and the in--phase N\'{e}el flux solution
was analysed at zero temperature.
The author found
that the one--dimensional limit is singular,
i.e. the staggered magnetization
$m\ne 0$ ($2m=0.513$)
when $J_{\perp}\to +0$
(although we know from exact results
that for the antiferromagnetic chain $m=0$).
In the other limiting case $J=J_{\perp}$
the theory based on the fermionization procedure
yields $2m=0.778$
(the spin wave result is $2m=0.6$;
more accurate calculations predict $m=0.3074$
(see Ref. \onlinecite{019})).
The result of Ref. \onlinecite{008}
stays somewhat separately
in the estimate of the value of $\frac{J_{\perp}}{J}$
at which the staggered magnetization appears.
Different theories predict
$\frac{J_{\perp}}{J}$
from $0$ to $0.2$
(for details see Ref. \onlinecite{019}).
Ref. \onlinecite{008}
predicts $\frac{J_{\perp}}{J}=0$,
moreover $2m$ jumps from zero to $0.513$
for any infinitesimally small $\frac{J_{\perp}}{J}$.

The antiferromagnetic Heisenberg model on a ladder
within the frames of the in--phase N\'{e}el phase solution
(the ground state energy,
the singlet--triplet energy gap)
was discussed in Ref. \onlinecite{009}.
The effects of the interladder interaction and magnetic field
on the susceptibility
at nonzero temperatures
were studied in Ref. \onlinecite{011}.

A consideration of the spin--Peierls state under magnetic field
was reported in Ref. \onlinecite{010}.
A study of the stepped spin--Peierls transition
for the quasi--one--dimensional $XY$ and Heisenberg models
using the 2D Jordan--Wigner transformation
was reported in Refs. \onlinecite{012,013}.
The 2D Jordan--Wigner transformation
was applied
for a study of the zero temperature spin--Peierls transition
for the quasi--one--dimensional $XY$ and Heisenberg systems
in Ref. \onlinecite{014}.
In particular,
the phase diagram between the dimerised and uniform states
in the parameter space of interchain interaction 
and spin--lattice coupling was constructed.

Many more problems may be considered
within the frames of the 2D Jordan--Wigner fermionization approach.
Probably,
the 2D Jordan--Wigner transformation
should be of more use
for the 2D spin--$\frac{1}{2}$ $XY$ models
since for such models no further approximations
(except the mean--field--like treatment of the phase factors)
are required.
Besides, a more sophisticated treatment of the phase factors
is desirable.

\vspace{5mm}

The author acknowledges the kind hospitality
of the Max Planck Institute for the Physics of Complex Systems, Dresden
in the end of 1999
when the paper was launched.
He thanks J. Richter (Magdeburg University)
for the kind hospitality
in the summer of 2000
when the main part of the paper
was prepared and discussed.
Special thanks go to 
T. Krokhmalskii, 
T. Verkholyak,
O. Zaburannyi 
and V. Derzhko
for helpful comments and suggestions.


\begin{thebibliography}{}

\bibitem{001}
  E. Lieb, T. Schultz, and D. Mattis,
  Ann. Phys. (N.Y.) {\bf 16,} 407 (1961);\\
  see also
  S. Katsura, Phys. Rev. {\bf 127,} 1508 (1962).

\bibitem{002}
  T. D. Schultz, D. C. Mattis, and E. H. Lieb,
  Rev. Mod. Phys. {\bf 36,} 856 (1964).

\bibitem{003}
  I. Affleck,
  Field theory methods and quantum critical phenomena.
  In:
  Fields, Strings and Critical Phenomena,
  E. Br\'{e}zin and J. Zinn--Justin, eds.,
  Elsevier Science Publishers B. V., 1989, p.563--640.

\bibitem{004}
  E. Fradkin,
  Phys. Rev. Lett. {\bf 63,} 322 (1989).

\bibitem{005}
  Y. R. Wang,
  Phys. Rev. B {\bf 43,} 3786 (1991).

\bibitem{006}
  Y. R. Wang,
  Phys. Rev. B {\bf 43,} 13774 (1991).

\bibitem{007}
  Y. R. Wang,
  Phys. Rev. {\bf 46,} 151 (1992).

\bibitem{008}
  M. Azzouz,
  Phys. Rev. B {\bf 48,} 6136 (1993).

\bibitem{009}
  M. Azzouz, L. Chen, and S. Moukouri,
  Phys. Rev. B {\bf 50,} 6233 (1994).

\bibitem{010}
  M. Azzouz and C. Bourbonnais,
  Phys. Rev. B {\bf 53,} 5090 (1996).

\bibitem{011}
  M. Azzouz, B. Dumoulin, and A. Benyoussef,
  Phys. Rev. B {\bf 55,} R11957 (1997).

\bibitem{012}
  Y. Ji, J. Qi, J.--X. Li, and C.--D. Gong,
  J. Phys.: Condens. Matter {\bf 9,} 2259 (1997).

\bibitem{013}
  X.--J. Fan and C.--D. Gong,
  Eur. Phys. J. B {\bf 7,} 233 (1999).

\bibitem{014}
  Q. Yuan, Y. Zhang, and H. Chen,
  cond--mat/9911119.

\bibitem{015}
  L. Huerta and J. Zanelli,
  Phys. Rev. Lett. {\bf 71,} 3622 (1993).

\bibitem{016}
  B. Bock and M. Azzouz,
  cond--mat/0007261.

\bibitem{017}
  V. Kalmeyer and R. B. Laughlin,
  Phys. Rev. Lett. {\bf 59,} 2095 (1987).

\bibitem{018}
  R. B. Laughlin,
  Phys. Rev. Lett. {\bf 60,} 2677 (1988).

\bibitem{019}
  D. Ihle, C. Schindelin, A. Wei\ss e, and H. Fehske,
  Phys. Rev. B {\bf 60,} 9240 (1999).

\end{thebibliography}
\end{document}